\documentclass[reprint,aps,prl,superscriptaddress,noeprint]{revtex4-2}

\usepackage[normalem]{ulem}
\usepackage{siunitx}
\usepackage{amsmath}
\usepackage{amsbsy}
\usepackage{amssymb} 
\usepackage{graphicx}
\usepackage{float}
\usepackage{bm}
\usepackage[caption=false]{subfig}
\usepackage{hyperref}
\hypersetup{
	colorlinks=true,
	linkcolor=blue,
	citecolor=green,
	filecolor=magenta,
	urlcolor=blue
}
\usepackage{siunitx}
\usepackage[dvipsnames]{xcolor}
\usepackage{booktabs}
\usepackage{multirow}

\begin{document}
\title{Engineering One Axis Twisting via a Dissipative Berry Phase Using Strong Symmetries}	

\author{Jeremy T. Young}
\email[Corresponding author: ]{j.t.young@uva.nl}
\affiliation{Institute of Physics, University of Amsterdam, 1098 XH Amsterdam, the Netherlands}
\affiliation{JILA, University of Colorado and National Institute of Standards and Technology, and Department of Physics, University of Colorado, Boulder, Colorado 80309, USA}
\affiliation{Center for Theory of Quantum Matter, University of Colorado, Boulder, Colorado 80309, USA}

\author{Edwin Chaparro}
\affiliation{JILA, University of Colorado and National Institute of Standards and Technology, and Department of Physics, University of Colorado, Boulder, Colorado 80309, USA}
\affiliation{Center for Theory of Quantum Matter, University of Colorado, Boulder, Colorado 80309, USA}

\author{Asier Pi\~{n}eiro Orioli}
\affiliation{JILA, University of Colorado and National Institute of Standards and Technology, and Department of Physics, University of Colorado, Boulder, Colorado 80309, USA}
\affiliation{Center for Theory of Quantum Matter, University of Colorado, Boulder, Colorado 80309, USA} 

\author{James K. Thompson}
\affiliation{JILA, University of Colorado and National Institute of Standards and Technology, and Department of Physics, University of Colorado, Boulder, Colorado 80309, USA}

\author{Ana Maria Rey}
\affiliation{JILA, University of Colorado and National Institute of Standards and Technology, and Department of Physics, University of Colorado, Boulder, Colorado 80309, USA}
\affiliation{Center for Theory of Quantum Matter, University of Colorado, Boulder, Colorado 80309, USA}

\date{\today}

\begin{abstract}
We show how a driven-dissipative cavity coupled to a collective ensemble of atoms can dynamically generate metrologically useful spin-squeezed states. In contrast to other dissipative approaches, we do not rely on complex engineered dissipation or input states, nor do we require tuning the system to a critical point. Instead, we utilize a strong symmetry, a special type of symmetry that can occur in open quantum systems and emerges naturally in systems with collective dissipation, such as superradiance. This symmetry  preserves coherence and allows for the accumulation of an atom number-dependent Berry phase which in turn creates spin-squeezed states via emergent one-axis twisting dynamics. This work shows that it is possible to generate entanglement in an atom-cavity resonant regime with macroscopic optical excitations of the system, going beyond the typical dispersive regime with negligible optical excitations often utilized in current cavity-QED experiments. 
\end{abstract}


\maketitle

An important goal in quantum metrology is to develop methods for generating quantum entanglement that reduce measurement uncertainty beyond the standard quantum limit, a fundamental bound  which defines the maximum precision of phase estimation achievable with uncorrelated particles. 
One of the most common approaches to generating metrologically-useful entanglement is via cavity-QED systems, where the cavity photons mediate highly-collective spin interactions between $N$ particles, whose resulting collective interactions are ideal for preparing spin-squeezed states \cite{Kitagawa1993, Wineland1992, Wineland1994, Pezze2018}. 

\begin{figure}[h!]
\vspace{.12cm}
    \centering
    \includegraphics[scale=1]{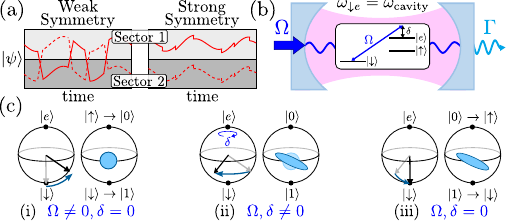}
    \caption{
    (a) Qualitative illustration of symmetry breaking for two quantum trajectories of the wavefunction (solid, dashed). Weak: Quantum jumps move between two symmetry sectors (light, dark); symmetry breaking conserves symmetry on average by conserving the probability to be in a given sector \cite{traj}.
    Strong: Symmetry conserved for each trajectory; symmetry breaking ensures the environment cannot distinguish which sector a given trajectory corresponds to, preserving coherences between sectors.
    (b) An array of three level  atoms ($|{\uparrow}\rangle,|{\downarrow}\rangle,|e\rangle) $ are loaded in a driven  optical cavity  resonant with the  $|{\downarrow}\rangle-|e\rangle$ transition. 
    (c) Schematic of squeezing generation, see text for details. 
    \label{fig:schematic}
    }
\end{figure}

Simultaneously, extensive efforts have been devoted to the storage and protection of quantum information, including topologically-protected ground-state manifolds \cite{Kitaev2003,Nayak2008} and quantum error correction \cite{Shor1995,Terhal2015}. 
These have been extended to open quantum systems in the form of passive error correction \cite{Lidar1998,Paz1998,Barnes2000,Sarovar2005,Mirrahimi2014,Cohen2014,Moloney2015,Kapit2015,Kapit2016,Puri2017,Reiter2017,Lihm2018,Lieu2022}, which was recently shown \cite{Lieu2020, Liu2023} to be a consequence of spontaneous breaking of a ``strong symmetry'' \cite{Buca2012}, a symmetry whose conservation laws are satisfied at the single-trajectory level, which generally support multiple steady states as a result \cite{Buca2012,suppcite,Albert2014, Manzano2018,SanchezMunoz2019,Dutta2020,Minganti2023,Li2023,Zhang2024}, as opposed to ``weak'' \cite{Buca2012} or ``average'' \cite{Ma2023a, Ma2023b} symmetries which only exhibit conservation laws on average in the density matrix (see Fig.~\ref{fig:schematic}a). 
Formally, a strong symmetry is distinguished by the fact that it is an explicit symmetry of the jump operators, rather than a symmetry of the Lindblad superoperators as in the case of a weak symmetry,  ensuring the symmetry's conservation under quantum jumps.

We present a protocol to dynamically generate spin squeezing in a driven-dissipative optical cavity via a Berry phase using a strong symmetry. 
We consider a spin-1/2 system composed of long-lived states $|{\uparrow}\rangle,|{\downarrow}\rangle$ (e.g., clock/ground states). In its simplest implementation, the system is initialized in an equal superposition of $|{\uparrow}\rangle,|{\downarrow}\rangle$ and an external laser is used to drive $|{\downarrow}\rangle$ to an excited state $|e\rangle$, which is resonant with the cavity (see Fig.~\ref{fig:schematic}b).  Here, the strong symmetry corresponds to global phase rotations on the $|{\downarrow}\rangle, |e\rangle$ states, and the combined number of atoms in the two states is conserved, which emerges because $|{\uparrow}\rangle$ is not coupled to the other states.
The protocol consists of three steps (see Fig.~\ref{fig:schematic}c). (i) A resonant drive is turned on, and the interplay between coherent drive and collective dissipation brings the system to a steady state which defines an effective qubit state $|1\rangle$, composed of a superposition of $|{\downarrow}\rangle, |e\rangle$, and $|0\rangle \equiv |{\uparrow}\rangle$.
(ii) The drive is then detuned from resonance so the phase of the drive changes in time, leading to a corresponding change in the phase of the steady state, simultaneously resulting in the accumulation of a global phase on $|{\downarrow} \rangle, |e\rangle$ (which form $|1\rangle$) relative to $|{\uparrow}\rangle$.
(iii) After the drive is turned off, the accumulated phase is preserved as $|e\rangle$ relaxes to $|{\downarrow}\rangle$, i.e., $|1\rangle \to |{\downarrow}\rangle$, due to the strong symmetry preserving relative phases between ($|{\downarrow}\rangle, |e\rangle$) and $|{\uparrow}\rangle$.

The accumulated phase on $|{\downarrow}\rangle$ is thus a dissipative Berry phase $\phi_B$ defined by the closed loop formed by the steady-state Bloch vector. We utilize a gauge where the accumulation of $\phi_B$ only occurs during (ii), so we focus primarily on the dynamics of this step unless otherwise noted. 
Because $\phi_B$ is sensitive to variations in the number of initial $|{\downarrow}\rangle$ atoms, the Berry phase generates effective one-axis twisting (OAT) dynamics, a paradigmatic fully connected Ising model ($\hat S_z^2$) \cite{Kitagawa1993}, 
which shears an initial coherent spin state distribution into an entangled state with scalable (in $N$) spin squeezing.
The strong symmetry helps preserve the accumulated phase throughout the dynamics and allows for the turn on/off of the drive to be done quickly \cite{suppcite}. Steps (i,ii) can be combined due to the rapid relaxation to the steady state $N \Gamma \gg \delta$. 

We show that our protocol achieves scaling comparable to or even better than state-of-the-art spin squeezing generation protocols in the presence of realistic sources of decoherence. 
Given that our protocol is not limited to the weak-excitation  regime of many coherent \cite{Schleier-Smith2010, Borregaard2017, Pedrozo-Penafiel2020, Colombo2022, Borregaard2017a, Luo2024, Leroux2010, Braverman2019, Barberena2023b}, measurement-based \cite{Schleier-Smith2010a, Cox2016, Hosten2016, Thomsen2002, Chen2014,Bao2020, Bowden2020, Robinson2022, Greve2022, Barberena2023b}, and dissipative \cite{DallaTorre2013} approaches, it enables much faster squeezing generation timescales, which can reduce the detrimental effect of other types of decoherence processes beyond those intrinsic to the cavity setup. Unlike most dissipative schemes which achieve spin squeezing in the steady state of a driven cavity near criticality \cite{DallaTorre2013, Lee2014a, Fernandez-Lorenzo2017, Groszkowski2022,Pavlov2023, Barberena2024,Gonzalez-Tudela2013, Wolfe2014, Somech2024}, our protocol dynamically generates squeezing away from criticality and stored in metrologically-relevant states (e.g., a clock transition) after the drive is turned off. Finally, unlike measurement-based approaches \cite{Thomsen2002,Schleier-Smith2010a, Chen2014, Cox2016, Hosten2016, Bao2020, Bowden2020, Robinson2022, Greve2022, Barberena2023b}, our protocol is deterministic and not limited by detection efficiency.

\emph{Model}.---We consider a system composed of a single driven lossy cavity mode $\hat a$ that resonantly interacts with $|{\downarrow} \rangle$ and an excited state $|e \rangle$ with single-photon Rabi frequency $g_c$ and power decay rate $\kappa$ (see Fig.~\ref{fig:schematic}b). We focus on the bad cavity limit $\kappa \gg \sqrt{N} g_c$,
so we can adiabatically eliminate the cavity \cite{suppcite}.
The spin dynamics of the system are described by the master equation
\begin{subequations}
\begin{equation}
\label{eq:spin_model}
\partial_t \hat\rho = - i [\hat H_{B,\delta} ,\hat\rho] + \Gamma \left( \hat l \hat\rho \hat l^\dagger - \frac{1}{2} \{\hat l^\dagger \hat l, \hat\rho \} \right),
\end{equation}
\begin{equation}
    \hat H_B = \frac{\Omega \hat J^+ + \Omega^* \hat J^-}{2}, \quad \hat H_\delta = |\Omega| \hat J_x - \delta \hat N_e,
\end{equation}
\end{subequations}
for an effective Rabi frequency $\Omega$ due to the cavity drive. The Hamiltonians correspond to the rotating frame of the atomic $|{\downarrow}\rangle \to |e\rangle$ transition frequency ($\hat H_B$) or the drive frequency ($\hat H_\delta$), where $\hat J^- \equiv \sum_i |{\downarrow}_i\rangle \langle e_i|$ defines an effective two-level collective operator in the relevant frame, $\hat N_e \equiv \sum_i |e_i\rangle \langle e_i|$, and $\Gamma=4 g_c^2/\kappa$ is the collective decay rate with Lindblad jump operator $\hat l = \hat J^-$. 

The two equivalent versions of the Hamiltonian support two different perspectives for the dynamics. $\hat H_B$ supports a geometrical picture where slow  variations in $\Omega$, via
$\Omega = |\Omega| e^{i \delta t}$,
result in the accumulation of a Berry phase. Alternatively, the time-dependence is eliminated in the rotating frame of the drive for $\hat H_\delta$, where $\delta$ physically corresponds to the detuning of the drive (see Fig.~\ref{fig:schematic}b). The latter further helps to clarify non-adiabatic effects in the dynamics, including decoherence. Throughout, we will use $\hat H_\delta$ for its convenience for calculations.

\emph{Mean field theory}.---The two-level piece $(|{\downarrow}\rangle-|e\rangle)$ of the above model has previously been studied in several contexts \cite{Drummond1978, Puri1980, Walls1980, Kilin1980, Iemini2018, Hannukainen2018, Barberena2019, Link2019, Somech2024, Barberena2024,Gonzalez-Tudela2013, Wolfe2014}, so we summarize key features assuming $N_J$ total atoms within this manifold. First, we recast the spin operators in terms of Schwinger bosons $\hat d, \hat e$, bosonic operators that annihilate atoms in the $|{\downarrow}\rangle$ and $|e\rangle$
states. In terms of these bosonic operators,  $\hat{J}^- \equiv \hat d^\dagger \hat e ,\hat{J}_z \equiv (\hat e^\dagger \hat e - \hat d^\dagger \hat d)/2$. In the large $N_J$ limit, the mean-field equations are
\begin{subequations}
    \begin{gather}
        \partial_t d = - i \frac{\Omega}{2} e + i \omega_B[N_J] d + \frac{\Gamma}{2} |e|^2 d,\\
        \partial_t e = - i \frac{\Omega}{2} d + i (\delta + \omega_B[N_J ]) e - \frac{\Gamma}{2} |d|^2 e,
    \end{gather}
\end{subequations}
where $d \equiv e^{i \omega_B[N_J] t} \langle \hat d\rangle, e \equiv e^{i \omega_B[N_J] t} \langle \hat e \rangle$. Here, we have gone to a rotating frame defined by $\omega_B[N_J]$, which is a dynamical phase that emerges due to Berry phase accumulation. It is determined by requiring that the above differential equations reach a steady state in the rotating frame \cite{suppcite}
\begin{equation}
    \omega_B[N_J ] = \frac{\delta}{2 \cos \theta_J[N_J]} - \frac{\delta}{2},
\end{equation}
where $\cos \theta_J[N_J] \equiv - 2\langle \hat J_z \rangle/N_J$. Physically, this describes the accumulation of a single-particle global phase on $|{\downarrow}\rangle, |e\rangle$.

When $\delta = 0$, the two-level system features spin-polarized and mixed steady-state phases separated by a  critical point  at $\Omega_c[N_J] = \frac{N_J \Gamma}{2}$, which was recently observed for the first time \cite{Song2024}.
When $\Omega > \Omega_c$, the system is in a mixed phase which exhibits zero steady state inversion ($\langle \hat J_z \rangle = 0$). In this phase, the drive overwhelms the collective dissipation, leading to modified Rabi cycles whose competition with the collective dissipation causes the spins to decohere. A Berry phase cannot be defined in such conditions.

\begin{figure}
    \centering
    \includegraphics{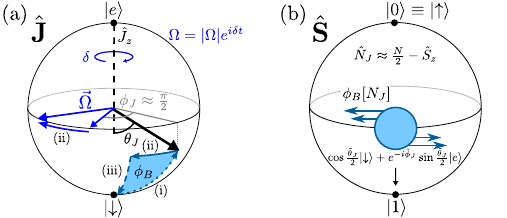}
    \caption{Bloch sphere description. (a) Bloch sphere dynamics for two-level model $\hat J$. The Berry phase accumulation process is shown for each step of the protocol. (b) Resulting dynamics on effective $\hat S$ Bloch sphere formed from the $|0\rangle, |1\rangle$ states during step (ii).}
    \label{fig:BlochSpheres}
\end{figure}

When $\Omega < \Omega_c$, the system is  in the spin-polarized phase. The steady state exhibits nonzero inversion and collective coherence due to the competition of the drive and dissipation, so there are no photons inside the cavity in this phase, $\langle \hat a \rangle = i \frac{g_c}{\kappa/2} (\langle \hat J^- \rangle + i \Omega/\Gamma) = 0$ \cite{suppcite}. For large $N_J$, the steady state  is  highly  pure and admits an excellent  mean-field description:
\begin{equation}
     |\psi_{N_J}\rangle_\text{ss}  \approx 
     \left(\cos \frac{\theta_J[N_J]}{2} |{\downarrow}\rangle + e^{-i \phi_J[N_J]} \sin \frac{\theta_J[N_J]}{2} |e \rangle\right)^{N_J},
\end{equation}
where $\cos \theta_J[N_J] = \sqrt{1 - |\Omega/\Omega_c[N_J]|^2}$, $\phi_J[N_J] = \pi/2$. When $0 < |\delta| \ll N_J \Gamma$, $\phi_J[N_J]$ is slightly perturbed, leading to small photon buildup in the cavity.  In this case, changes in the phase of $\Omega$ induce equivalent changes in $\phi_J$, corresponding to step (ii) in the protocol. The steady state in this regime is illustrated in Fig.~\ref{fig:BlochSpheres}a.

\emph{Strong symmetry}.---Now consider an initial superposition $|\psi(0)\rangle = ( |{\uparrow}\rangle/\sqrt{2} + e^{i \varphi} |{\downarrow}\rangle/\sqrt{2})^N$. Since the $|{\uparrow}\rangle$ state is not coupled to the $\{|{\downarrow}\rangle, |e\rangle \}$ states, the state in the spin-polarized phase during (ii) becomes
\begin{equation}
\label{eq:steadypsi}
 |\psi(t) \rangle \approx \sum_{N_J = 0}^N \sqrt{\binom{N}{N_J}} \frac{ e^{i N_J \varphi}}{2^N}  e^{i \phi_B[N_J,t]} \mathcal{S}|\psi_{N_J} \rangle_\text{ss} |{\uparrow}\rangle^{N_{\uparrow}}, 
\end{equation} 
where $N_{\uparrow} = N - N_J$, $\mathcal{S}$ symmetrizes the $\{|{\downarrow}\rangle, |e\rangle \}$ and $|{\uparrow}\rangle$ states, and $\phi_B[N_J,t]$ captures the $N_J$ and time dependence of the Berry phase accumulation, which reduces to $\omega_B t$ at the single-particle mean-field level. 

To understand the dynamics, we define a basis $|0\rangle \equiv |{\uparrow} \rangle, |1 \rangle \equiv \cos \frac{\tilde{\theta}_J}{2} |{\downarrow} \rangle + e^{-i \tilde{\phi}_J} \sin \frac{\tilde{\theta}_J}{2} |e\rangle$ with collective spin $\hat S$, where tildes denote mean-field values at $N_J = \langle \hat N_J \rangle = N/2$. In this basis, $ \mathcal{S}|\psi_{N_J} \rangle |{\uparrow}\rangle^{N_{\uparrow}}$ describe collective Dicke states of $\hat S$ with $m_S = \frac{N_{\uparrow} - N_J}{2}$, so $\hat S_z \approx \frac{N}{2} - \hat N_J$, and the evolution of $\phi_B[N_J,t]$ approximately corresponds to phase evolution for a specific $m_S$ state. Since $N_J$ for each term in the superposition is conserved, they
evolve independently, accumulating different Berry phases. This in turn leads to a shearing of the noise distribution as observed in standard OAT dynamics, which similarly preserves the total magnetization (see Fig.~\ref{fig:BlochSpheres}b).

The $\hat S$ picture is possible because in the spin-polarized phase, a ``strong'' symmetry \cite{Buca2012} undergoes spontaneous symmetry breaking.
Strong symmetries leave the Lindblad jump operators invariant, enforcing conserved quantities on an individual trajectory level \cite{Buca2012, SanchezMunoz2019, Lieu2020, Dutta2020,suppcite,Albert2014, Manzano2018,Zhang2024, Li2023,Liu2023,Ma2023a, Ma2023b, Minganti2023}. Here, the symmetry corresponds to phase rotations $|{\downarrow}_i\rangle, |e_i\rangle \to e^{i \varphi'} |{\downarrow}_i\rangle, e^{i \varphi'} |e_{i}\rangle$ for all $i$ and number conservation $\hat N_J \equiv \sum_i |{\downarrow}_{i} \rangle \langle {\downarrow}_{i}| + |e_{i} \rangle \langle e_{i}|$, where eigenvalues of $\hat N_J$ define the symmetry sectors. Conservation of $\hat N_J$ implies conservation of the $|{\uparrow}\rangle$ population, while strong-symmetry breaking preserves the coherences between terms with different $N_J$ \cite{Lieu2020,Liu2023}.

Physically, this means for any quantum trajectory in one sector (e.g., states with exactly $N_J$ atoms in $|{\downarrow}\rangle, |e\rangle$), equivalent trajectories exist for the other sectors ($N_J' \neq N_J$ atoms in $|{\downarrow}\rangle, |e\rangle$) and thus are indistinguishable from the perspective of the environment. For a weak symmetry, quantum jumps could take $N_J$ states to $N_J'$ states and vice-versa, and $\hat N_J$ could only be conserved on average (see Fig.~\ref{fig:schematic}a). Thus, the strong symmetry prevents the environment from projecting the system into a specific sector, thereby preserving the coherences between sectors. Formally, the steady-state space forms a decoherence-free subspace \cite{Lidar1998}, although strong symmetries can also result in more general noiseless subsystems \cite{Knill2000, Albert2014, Lieu2020}.

The  strong symmetry further ensures that the squeezing generated outside the $|{\uparrow}\rangle ,|{\downarrow} \rangle$ manifold is mapped back to $|{\uparrow}\rangle,|{\downarrow}\rangle$ by simply turning off the drive. 
All that occurs is $\mathcal{S}|\psi_{N_J} \rangle |{\uparrow}\rangle^{N_{\uparrow}} + \mathcal{S}|\psi_{N_J'} \rangle |{\uparrow}\rangle^{N_{\uparrow}'} \to \mathcal{S}|{\downarrow} \rangle^{N_J}|{\uparrow}\rangle^{N_{\uparrow}}+\mathcal{S}|{\downarrow} \rangle^{N_J'}|{\uparrow}\rangle^{N_{\uparrow}'}$ with no state mixing when $|N_J - N_J'| < \mathcal{O}(\sqrt{N})$  \cite{suppcite}, so the Berry phase evolution is retained. 
This allows us to consider dynamics with significant $|e\rangle$ excitations while nevertheless generating squeezing in the desired $|{\downarrow}\rangle, |{\uparrow}\rangle$ manifold.

{\it Squeezing dynamics}.---We are interested in deriving an effective Hamiltonian from the Berry phase $\hat {H}_{\text{eff}} |\Psi_{N_J} \rangle = (-\partial_t  \phi_B) |\Psi_{N_J}\rangle$, where $|\Psi_{N_J} \rangle \equiv \mathcal{S}|\psi_{N_J} \rangle_\text{ss} |{\uparrow}\rangle^{N_{\uparrow}}$.
We note that, for large $N_J$,  the  accumulated  Berry phase is that of a large spin in a magnetic field and  given by the subtended solid angle, $\frac{N_J}{2}(1- \cos \theta_J)$. Therefore,   $\partial_t \phi_B =  -\delta [N_J (1-\sqrt{1-\sin^2 \theta_J})]/2$. Since for small $\delta$, $\sin \theta_J \approx  \Omega/(N_J \Gamma/2)$, we incorporate quantum fluctuations in $\phi_B$ by replacing  $N_J \to \hat{N}_J = N/2 - \hat S_z$  and  $\sin \theta_J \to \sin \theta_J[\hat{N}_J]\approx \frac{N/2}{\hat N_J} \sin \tilde{\theta}_J$ in $\partial_t \phi_B$. 
Expanding in powers of $|\hat S_z/N|\ll 1$, $\hat {H}_{\text{eff}}$ becomes 
\begin{equation}
\label{eq:Heff}
     \hat H_{\text{eff}} \approx  -\tilde{\omega}_B \hat S_z + \check{\chi} \hat S_z^2 + \cdots, \quad \check{\chi} = - \frac{\delta \sin^2 \tilde{\theta}_J}{2 N \cos^3 \tilde{\theta}_J},
\end{equation} where we have dropped a constant term as a global phase. Here, we see that variations of the Berry phase with $N_J$ give rise to OAT dynamics and describe the mean-field rotating frame of $\tilde{\omega}_B$.

When $\delta \neq 0$, the cavity's steady state is no longer the vacuum. The finite  coherence of the cavity,  proportional to the modified jump operator $\hat l = \hat J^- + i \Omega/\Gamma \approx -i (\kappa/2 g_c) \langle \hat a \rangle$, thus   depends on  the population of the  various $N_J$ terms, which radiate at different rates,   leading to dephasing. 
By using its  corresponding  mean-field value at linear order in $\delta$, $ \langle \hat l \rangle = N_J \left(\frac{\delta}{2 \Omega} \sin \theta_J\tan \theta_J - i \frac{N_J \Gamma}{4 \Omega} \sin^2 \theta_J \right) + i \Omega/\Gamma $, and  replacing  $N_J$ and $\sin\theta_J$ with their operator values as above, we obtain  
\begin{equation}
\label{eq:l}
    \hat l\approx \frac{\delta}{\Gamma} \tan \tilde{\theta}_J + \frac{2 \delta  \sin \tilde{\theta}_J}{N \Gamma \cos^3 \tilde{\theta}_J } \hat S_z + \cdots .
\end{equation} 
The linear $\hat S_z$ term indicates the presence of an effective collective dephasing due to the environment's ability to slowly gain information about $\hat N_J$. The constant term does not affect the $\hat S$ dynamics. The ratio of the OAT and collective dephasing rates is $N \Gamma \cos^3 \tilde{\theta}_J/(8 \delta)$, so the effect of collective dephasing is reduced when the detuning is small, corresponding to the adiabatic limit where everything is determined by the Berry phase accumulation and there is minimal distinguishability between the different $N_J$ terms in the state. This scaling is consistent with recent results investigating the manipulation of these steady-state spaces \cite{Santos2024}. Further details for both derivations are presented in the End Matter.

To extend the analysis beyond the small $\delta$ limit, we utilize a generalized Holstein-Primakoff approximation, from which we find \cite{suppcite}
\begin{subequations}
\label{eq:AEHP}
    \begin{equation}
        \hat H_{\text{eff}} =  -\frac{\delta}{2N } \frac{N^2 \Gamma^2 \sin \tilde{\theta}_J \tan \tilde{\theta}_J}{N^2 \Gamma^2 \cos^2 \tilde{\theta}_J + 16 \delta^2 \sec^2 \tilde{\theta}_J} \hat S_z^2,
    \end{equation}
    \begin{equation}
        \hat l \approx 2  e^{-i \tilde{\phi}_J} \frac{\delta \tan \tilde{\theta}_J (i N \Gamma - 4 \delta \sec \tilde{\theta}_J ) }{N^2 \Gamma^2 \cos^2 \tilde{\theta}_J + 16 \delta^2 \sec^2 \tilde{\theta}_J} \hat S_z,
    \end{equation}
\end{subequations}
where the constant term has been dropped from the jump operator. We benchmark these equations with quantum trajectories in the End Matter.

Most typical squeezing protocols are restricted to the perturbative, weak-excitation regime where $|e\rangle$ is not strongly populated and can be adiabatically eliminated \cite{Schleier-Smith2010, Borregaard2017, Pedrozo-Penafiel2020, Colombo2022, Borregaard2017a, Luo2024, Leroux2010, Braverman2019, Barberena2023b,Mivehvar2021}. However, the dynamics become correspondingly slower due to the reduced effective atom-cavity interaction. The slower dynamics  in the weak-excitation limit can be seen  in our protocol via the fact that both the OAT and dephasing rates are proportional to $\sin^2 \tilde{\theta}_J$, and  thus they become slow for weak excitations since  $\sin^2 \tilde{\theta}_J \to 0$. Nevertheless, our protocol extends beyond this regime, allowing for a  significant acceleration of the atom-cavity  dynamics. 
Therefore, while the scaling of our protocol with $N$ is the same as other OAT approaches, our protocol enables a dramatic reduction in the timescales necessary to achieve the same amount of squeezing by going beyond the perturbative regime.

{\it Single particle decoherence}.---Depending on the choice of states, there are different types of effective single particle decoherence in $\hat{S}$. A fundamental source of decoherence is spontaneous emission, which enters the master equation via
$\gamma_{e\updownarrow} \sum_i \left(\hat \sigma_i^{\updownarrow e} \hat \rho \hat \sigma_i^{e \updownarrow} - \frac{1}{2} \{\hat \rho , \hat \sigma_i^{ee}\} \right)$, where $ \hat \sigma^{ab} \equiv |a \rangle \langle b|$,
describing single particle  decay from $|e \rangle$ to either $|{\updownarrow}\rangle =|{\uparrow}\rangle$ or $ |{\downarrow} \rangle$, each of which has different effects on the effective $\hat S$ Bloch sphere dynamics. We model spontaneous emission to $|{\uparrow}\rangle$ at rate $\gamma_{e {\uparrow}}$ as an effective spin flip from $|1\rangle$ to $|0 \rangle$ at rate $\gamma_- \equiv \gamma_{e{\uparrow}} \sin^2 \frac{\tilde{\theta}_J}{2}$, which is reduced from $\gamma_{e{\uparrow}}$ by the excited state fraction of $|1 \rangle$.

We treat spontaneous emission to $|{\downarrow} \rangle$ at rate $\gamma_{e \downarrow}$ as an effective dephasing process in the reduced $|0\rangle, |1\rangle$ basis, which enters the effective master equation for $\hat S$ as $\gamma_d \sum_i \left(\hat \sigma_i^{11} \hat \rho_S \hat \sigma_i^{11} - \frac{1}{2} \{\hat \rho_S , \hat \sigma_i^{11}\} \right)$
with $\gamma_d = \gamma_{e \downarrow} \sin^2 \frac{\tilde{\theta}_J}{2}$. However, given that in typical experiments there are other sources of effective single-particle dephasing, in the discussion below we  assume a $\tilde{\theta}_J$-independent rate. If $|{\downarrow}\rangle \equiv {^1S_0}, |e\rangle \equiv {^3 P_1}, |{\uparrow} \rangle \equiv {^3 P_0}$ for $^{87,88}$Sr, creating squeezing on a clock transition, then $\gamma_{e{\uparrow}} = 0$, and there are no effective spin flips. In contrast, effective single-particle dephasing is always present.

\begin{figure}
    \centering
    \includegraphics[scale=.29]{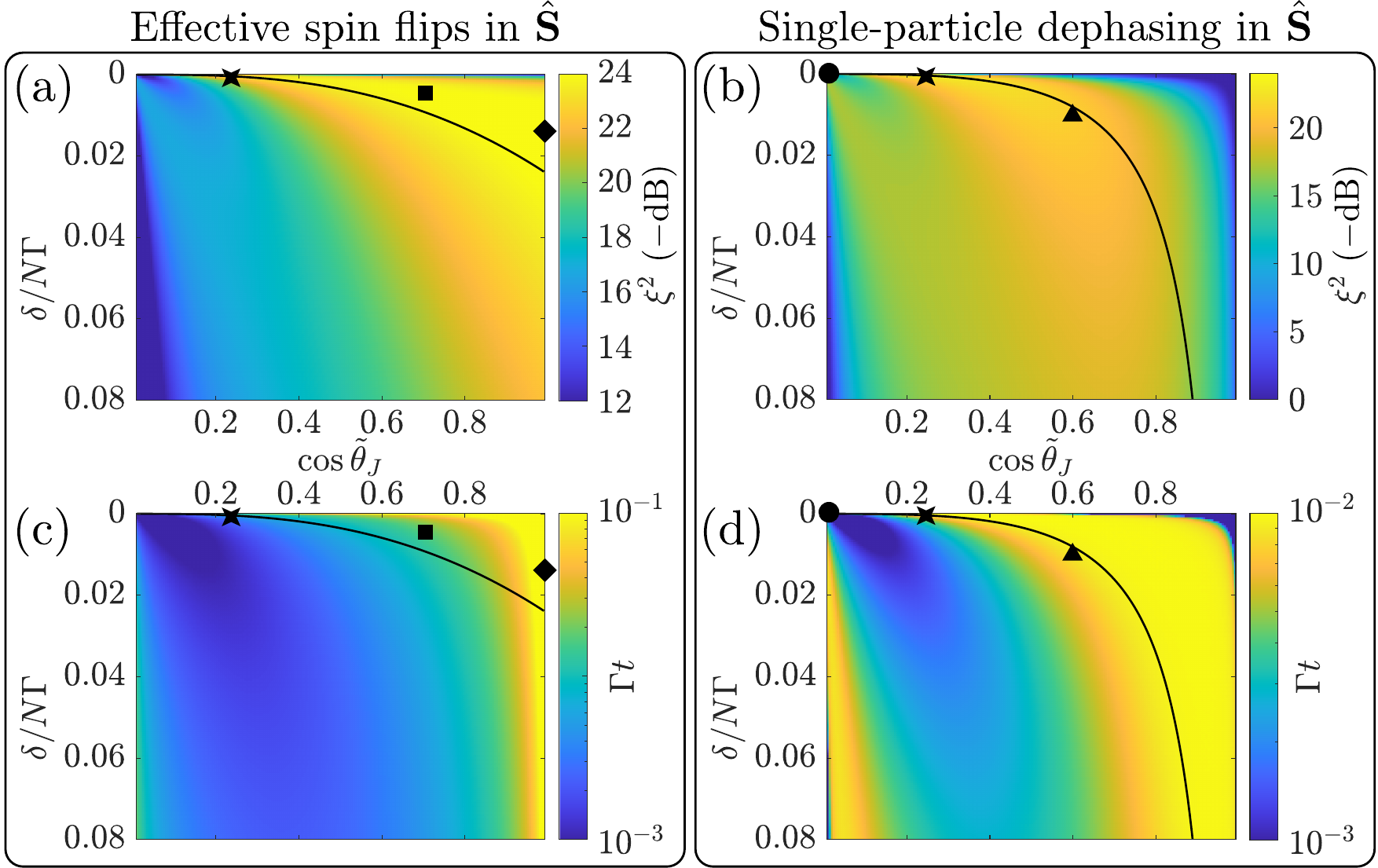}
    \caption{Optimal squeezing (top) and squeezing time (bottom) as a function of $\delta/N \Gamma, \cos \tilde{\theta}_J$ for $N=10^6$ using Eq.~(\ref{eq:AEHP}). (a,c) Only effective spin flips in $\hat{S}$: $\gamma_{-} = \Gamma \sin^2\frac{\tilde{\theta}_J}{2}, \gamma_d = 0$. (b,d) Only single-particle dephasing: $ \gamma_- = 0, \gamma_d = 100 \Gamma$. Black lines indicate the optimal $\delta$ for the most squeezing as a function of $\cos \tilde{\theta}_J$ using perturbative expansions. Shapes correspond to parameters discussed in main text.
    }
    \label{fig:AEHP}
\end{figure}

We quantify the squeezing via the Wineland squeezing parameter $\xi$ \cite{Wineland1992, Wineland1994}, which describes the reduction in phase measurement uncertainty beyond the standard quantum limit and is defined as $\xi^2 = \frac{N \text{min} (\Delta \hat{S}_{\perp})^2}{|\langle \hat{\mathbf{S}} \rangle|^2}$, where $\text{min}(\Delta \hat{S}_{\perp})^2$ denotes the minimum variance in directions perpendicular to $\langle \hat{\mathbf{S}} \rangle$. We analytically optimize the squeezing via an expansion of the $\hat S$ dynamics \cite{suppcite, Chu2021, Barberena2023b}
\begin{equation}
    \xi^2(t) \approx e^{\gamma_d t}\left(\frac{e^{\gamma_d t} + N \check{\Gamma} t}{N^2 \check{\chi}^2 t^2} + \frac{N^2 \check{\chi}^4 t^4}{6} + \frac{2}{3} \gamma_- t\right),
\end{equation}
where $\check \chi, \check \Gamma$ denote the OAT and collective dephasing rates, respectively. 
In Table \ref{tab:optsq}, we optimize the squeezing when one source of single-particle dephasing dominates.
As an example, in Fig.~\ref{fig:AEHP} we plot $\xi^2_{\text{opt}}$ and $t_{\text{opt}}$
for $N=10^6$, with either only effective spin flips or only effective single-particle dephasing, determined via an exact solution \cite{Chu2021} using Eq.~(\ref{eq:AEHP}) for the collective dynamics. 

\begin{table}
\centering
\def\arraystretch{1}
\setlength\tabcolsep{3.5mm}
\begin{tabular}{lcc}
\toprule
\toprule
   & $\xi^2_{\text{opt}}$ & $t_{\text{opt}}$ \\
\midrule
 Dephasing & $\propto \frac{1}{N^{2/3}}$ & $\min\left[\frac{3.04}{\sqrt{\Gamma /\gamma_d} N^{1/6}},1\right] \times \gamma_d^{-1}$\\
 Spin flips &  $ \frac{4\sqrt{2/3}}{\sqrt{N \Gamma/\gamma_{e{\uparrow}}}}$ & $\frac{\sqrt{6}}{ \sin^2 \frac{\tilde{\theta}_J}{2} \sqrt{N \Gamma/\gamma_{e{\uparrow}}}} \times \gamma_{e \uparrow}^{-1}$ \\
 \bottomrule
 \bottomrule
\end{tabular}
\caption{Optimal performance of squeezing protocol in the presence of single-particle dephasing. Similar scaling is realized by weak-excitation OAT protocols ($\sin^2 \frac{\tilde{\theta}_J}{2} \ll 1$) up to constant factors \cite{Schleier-Smith2010, Borregaard2017, Lewis-Swan2018, suppcite}. Scaling comparisons for other protocols and experimental values are reported in the supplement \cite{suppcite}. \label{tab:optsq}}
\end{table}

For spin flips ($\gamma_- = \Gamma \sin^2\frac{\tilde{\theta}_J}{2}, \gamma_d = 0$), the optimal squeezing is $24.4$ dB at $\delta = 0.014 N \Gamma$ in the limit $\sin^2\tilde{\theta}_J/2 \to 0$, i.e., $\Omega \to 0$ (diamond). However, the squeezing time scales grow like $1/\sin^{2} \frac{\tilde{\theta}_J}{2}$, so additional single-particle dephasing always shifts the optimal drive to stronger $\Omega$. 
Increasing the drive to $\cos \tilde{\theta}_J = 0.7 $, the squeezing at the optimal $\delta = 0.005 N \Gamma$ is $24.0$ dB in time $0.0163/\Gamma$ (square), which is 28 times faster than at the optimal detuning for $\cos \tilde{\theta}_J = 0.99$ (weak-excitation regime), so additional single-particle dephasing can be mitigated with minimal reduction in the squeezing.

For single-particle dephasing ($\gamma_- = 0, \gamma_d = 100 \Gamma$), the optimal squeezing is 23.1 dB in time $0.0085/\Gamma = 0.85/\gamma_d$, which occurs in the limit $\delta \to 0$ with $\cos^3 \tilde \theta_J \approx 33 \delta/N \Gamma$ (circle). However, in this limit finite-size effects which reduce the validity of Eq.~(\ref{eq:AEHP}) occur, so this limit cannot be realized. Nevertheless, moving away from criticality to $\cos \tilde{\theta}_J = 0.6$, which has optimal $\delta = 0.01 N \Gamma$ (triangle), there is again minimal reduction in the squeezing (21.3 dB) and minimal increase in the squeezing time ($0.009/\Gamma = 0.9/\gamma_d$).

When both types of decoherence are relevant, optimal squeezing occurs at intermediate drives. For $\gamma_- = \Gamma \sin^2\frac{\tilde{\theta}_J}{2}, \gamma_d = 100 \Gamma$, the optimal squeezing occurs at $\delta = 0.0004 N \Gamma, \cos \tilde{\theta}_J = 0.23$ with $20.8$ dB squeezing in time $0.0044/\Gamma = 0.44/\gamma_d$ (star).

\emph{Outlook}.---While we have focused on generating OAT dynamics, an important question is whether other types of dynamics can be generated, such as quantum non-demolition squeezing \cite{Schleier-Smith2010a, Hosten2016, Cox2016, Barberena2023b}.
Furthermore, an interesting direction is the generalization of our protocol to multilevel systems featuring a rich dark state structure \cite{PineiroOrioli2022, Lin2022, Fan2023}, which readily realize generalized forms of the strong symmetry considered here. Finally, investigations on  how predicted behaviors are affected by realistic experimental conditions  outside the  bad cavity limit or with inhomogeneous atom-light coupling are important for its implementation in optical clocks \cite{Robinson2022}.

\begin{acknowledgments}
    We thank B.~Sundar, D.~Barberena, S.~Lieu, and J.~Hur for helpful discussions and feedback. 
 This work is supported by the VBFF,  AFOSR grants FA9550-24-1-0179,  by the NSF JILA-PFC PHY-2317149, QLCI-OMA-2016244, by the U.S. Department of Energy, Office of Science, National Quantum Information Science Research Centers Quantum Systems Accelerator, and by NIST. J.T.Y.~was supported in part by the NWO Talent Programme (project number VI.Veni.222.312), which is (partly) financed by the Dutch Research Council (NWO).
\end{acknowledgments}

\bibliography{StrongSqueezing,Extra}

\begin{thebibliography}{86}
\bibitem [{tra()}]{traj}%
  \BibitemOpen
  \href@noop {} {\bibinfo {title} {{While we illustrate this as a pair of
  equivalent trajectories which exhibit an emergent detailed balance, the
  averaging can be much more complicated.}}}\BibitemShut {Stop}%
\end{thebibliography}

\appendix

\onecolumngrid
\newpage
\section{\large{End Matter}}

\twocolumngrid

\emph{Appendix A: Berry phase Hamiltonian}.---In this section, we show how the combination of the Berry phase with $N_J$ fluctuations gives rise to the effective Hamiltonian in the main text for adiabatic $\delta$. 
In the limit of large $N$, the steady state in the spin-polarized phase becomes a pure state that is approximately an eigenstate of $\hat J^-$ with eigenvalue $\alpha = j o e^{i \phi}$, where $j \equiv N_J/2$ and $o \equiv |\Omega/\Omega_c| = |\Omega|/(j \sqrt{\Gamma^2 + 4 \chi^2})$, where we have included collective elastic interactions $\chi \hat J^+ \hat J^-$ in the Hamiltonian \cite{Somech2024}. The eigenvalue phase $\phi$ is determined by the phase of $\Omega$ and the value of $\chi/\Gamma$ but is otherwise unimportant in determining the Berry phase of the steady state, although $\chi$ will modify the Liouvillian spectrum and its eigenstates. These eigenstates become exact exponentially fast with increasing $N_J$ provided $\langle \hat J_z \rangle = \cos \theta_J \gtrsim N_J^{-1/3}$, i.e., the system is not too close to the critical point where finite-size effects play a strong role. 
 
The corresponding eigenstate can be expressed \cite{Somech2024}
\begin{subequations}
\begin{equation}
    |\alpha \rangle = \sum_m a_m |j,m \rangle, \quad
    a_m = a_{-j}e^{f_m} e^{-i (N_J/2 + m) \phi}, 
\end{equation}
\begin{equation}
\begin{aligned}
    f_m ~&= (m+N_J/2) \log (o N_J/2) ~- \\ & \sum_{k = -N_J/2}^{m-1} \frac{1}{2} \log \left[\left(\frac{N_J}{2}+1\right) \frac{N_J}{2}-(k+1)k\right],
\end{aligned}
\end{equation}
\end{subequations}
where $|N_J/2,m\rangle$ are collective Dicke states and $a_{-j}$ is determined by normalizing the state. We select the gauge by taking $a_{-j}$ to be independent of $\phi$, corresponding to $\phi$ information being contained in $|e\rangle$, consistent with the rotating frame used for $\hat H_\delta$. Variations in the phase of $\Omega$ correspond to variations in $\phi$. Thus, the Berry connection in the $\phi$ direction is 
\begin{equation}
\begin{aligned}
    \mathcal{A}_\phi &= \frac{1}{r \sin \theta_J} i \langle \alpha| \nabla_\phi|\alpha \rangle = \frac{1}{r \sin \theta_J} \sum_m |a_m|^2 \left( \frac{N_J}{2} + m \right) \\
    &= \frac{\langle \hat J_z \rangle}{r \sin \theta_J} = \frac{N_J(1- \cos \theta_J)}{2 r \sin \theta_J} + \mathcal{O}(1),
\end{aligned}
\end{equation}
where we parametrize the variation according to the steady state Bloch vector, and the Berry connection is the same as for $N_J$ non-interacting atoms pointing in the direction of the steady state up to higher-order corrections in $1/N_J$. This applies to $\mathcal{A}_{\theta_J}, \mathcal{A}_r$ as well, so only $\mathcal{A}_\phi$ is relevant for the Berry phase in the gauge we have chosen. As a result, the phase accumulation rate of our protocol when $\delta$ is applied is 
\begin{equation}
\begin{aligned}
    \partial_t \phi_B[N_J] &= \delta \frac{N_J (1-\cos \theta_J)}{2}  + \mathcal{O}(1)\\
    &= \delta \frac{N_J (1-\sqrt{1 - o^2})}{2}+ \mathcal{O}(1),
\end{aligned}
\end{equation}
where we have utilized the fact that $\cos \theta_J = \sqrt{1-o^2} + \mathcal{O}(1/N_J)$ for adiabatic $\delta$.

As in the main text, we proceed to include fluctuations in $N_J$ to identify the effective dynamics. As such, we take
\begin{equation}
    N_J \to \frac{N}{2} - \hat S_z, \qquad o \to \frac{N/2}{N/2-\hat S_z} \tilde{o},
\end{equation}
where we have expressed $o$ in terms of the mean-field value $\tilde{o}$ at $\langle \hat N_J \rangle = N/2$. We expand in powers of $\hat S_z/N$, leading to
\begin{equation}
\begin{aligned}
    \partial_t \phi_B = \tilde{\omega}_B \hat S_z + \frac{\delta \sin^2 \tilde{\theta}_J}{2 N \cos^3 \tilde{\theta}_J} \hat S_z^2 + \cdots,
    \end{aligned}
\end{equation}
where $\tilde{o} \equiv \sin \tilde{\theta}_J$ and we have dropped an unimportant constant term. This describes the phase evolution of each $N_J$ term in the superposition, Eq.~(\ref{eq:steadypsi}), and hence the effective Hamiltonian $\hat H_\text{eff} \equiv -\partial_t \phi_B$.
Thus, we see that $\omega_B$ corresponds to the rotations in phase on the $\hat S$ Bloch sphere, which is removed in an appropriate rotating frame. We can include non-adiabatic corrections by incorporating the effect of $\delta$ on $\cos \theta_J$.

\emph{Appendix B: Collective dephasing}.---In this section, we identify the strength of collective dephasing in the adiabatic $\delta$ limit in a similar fashion. This is accomplished by considering how the emission of light is modified in the presence of a detuning. To understand this, we recast our dynamics as
\begin{equation}
    \hat H = -\delta \hat N_e, \qquad \hat l = \hat J^- + i \Omega/\Gamma,
\end{equation}
where the Rabi frequency has been absorbed into the jump operator. This further clarifies how in the resonant model ($\delta = 0$) the steady state is an eigenvector of $\hat J^-$: the corresponding eigenvector is a dark state of the modified jump operator. As discussed in the main text, this shifted jump operator is proportional to the steady-state cavity coherence and thus captures the distinguishability of different $N_J$ terms by the environment.

The role of $N_J$ fluctuations in the detuning Hamiltonian are already captured in the previous analysis of the Berry phase since $\langle \hat N_e \rangle = N_J(1- \cos \theta_J)/2$, so we focus on the remaining jump operator. Here, we utilize the mean-field values at linear order in $\delta$, particularly $\sin \theta_J = \frac{\Omega}{N_J \Gamma/2}$,
\begin{equation}
\begin{aligned}
    \langle \hat l \rangle &= N_J \sin^2 \theta_J\left(\frac{\delta}{2 \Omega} \sec \theta_J - i \frac{N_J \Gamma}{4 \Omega} \right) + i \Omega/\Gamma \\
    &= \frac{\delta}{\Gamma} \frac{o}{\sqrt{1-o^2}}\\
    & = \frac{\delta}{N \Gamma} \left(N \tan \tilde{\theta}_J + \frac{2 \sin \tilde{\theta}_J}{ \cos^3 \tilde{\theta}_J } \hat S_z + \cdots\right) ,
    \label{eq:flucdeph}
\end{aligned}
\end{equation}
where the last step has utilized the same approach to incorporating $N_J$ fluctuations as for the Berry phase. Thus, we see there that the steady state is no longer a dark state, leading to $\hat S_z$-dependent emission of the light. The linear $\hat S_z$ term defines an effective collective dephasing due to the environment's ability to slowly gain information about $N_J$, analogous to a quantum non-demolition measurement \cite{Schleier-Smith2010a, Hosten2016, Cox2016, Barberena2023b}. The constant term has no effect on the dynamics and can be dropped.

\emph{Appendix C: Benchmarking squeezing dynamics}.---In this section, we benchmark our analytic expressions for the OAT and collective dephasing rates with exact numerics. For the exact numerics, we utilize a quantum trajectory approach. While the whole Hilbert space is $\mathcal{O}(N^3)$, we utilize two simplifications. First, the strong symmetry allows us to express $\hat J$ in terms of independent operators for each $N_J$ sector, reducing the size of the operators we consider. Second, for the initial equal superposition of $|{\downarrow}\rangle$ and $|{\uparrow}\rangle$, the distribution of $N_J$ is a binomial distribution centered around $N_J = N/2$. Therefore, we restrict the range of $N_J$ in the initial state to $N/2 \pm dN$. Additionally, we utilize the jump operator $\hat l = \hat J^- + i \Omega/\Gamma$ to best reflect the actual photon loss, which also minimizes the frequency of jumps at the steady state.

To calculate the spin squeezing in the system in the middle of the protocol, we consider a generalization of the Wineland squeezing parameter. Taking $|c \rangle$ to be the single-particle state associated with the Bloch vector and identifying the remaining orthogonal states $|j\rangle, |s\rangle$ states, we capture the fluctuations perpendicular to the Bloch vector via the operators
\begin{subequations}
    \begin{equation}
         \sum_i \frac{|c_i \rangle \langle j_i| + |j_i \rangle \langle c_i|}{2}, \qquad
         \sum_i \frac{|c_i \rangle \langle j_i| - |j_i \rangle \langle c_i|}{2i},
    \end{equation}
    \begin{equation}
        \sum_i \frac{|c_i \rangle \langle s_i| + |s_i \rangle \langle c_i|}{2}, \qquad
        \sum_i \frac{|c_i \rangle \langle s_i| - |s_i \rangle \langle c_i|}{2i},
    \end{equation}
\end{subequations}
which are analogous to a pair of fluctuations with $x$ and $y$ components. With a proper choice of $|j\rangle,|s\rangle$, these can be understood as fluctuations on the $\hat J, \hat S$ Bloch spheres, respectively. We define an associated covariance matrix $\mathcal{C}$, from which the generalized (anti-)squeezing is given by eigenvalues of
\begin{equation}
    \frac{ N \mathcal{C}}{\langle \hat N_c/2\rangle^2},
\end{equation}
where $\hat N_c/2 \equiv \sum_i |c_i \rangle \langle c_i|/2$ defines the multilevel Bloch vector length. 

\begin{figure}[h]
    \centering
    \includegraphics[scale=0.5]{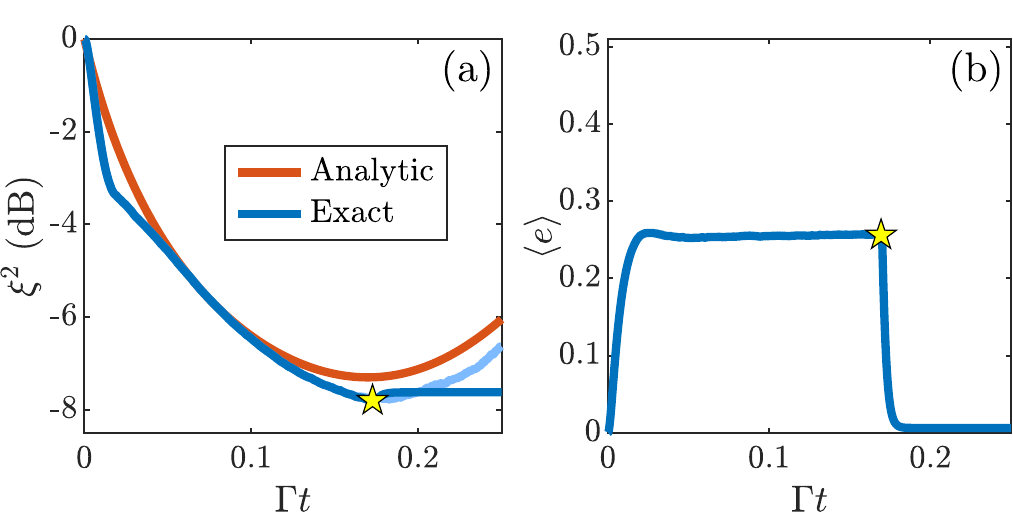}
    \caption{(a) Comparison of squeezing dynamics for exact quantum trajectories (blue) to analytics (red) for $\delta = 0.05 N \Gamma, \cos \tilde{\theta}_J \approx 0.5$.  For the dark blue line, the drive is turned off at $\Gamma t = 0.17$ (star). (b) Excited state fraction $\langle e \rangle$ dynamics during the protocol.}
    \label{fig:benchmark}
\end{figure}

In Fig.~\ref{fig:benchmark}, we compare the exact dynamics of the squeezing to Eq.~(\ref{eq:AEHP}) for $N=1000$, $dN = 80$, $ \delta = 0.05 N \Gamma, \Omega = 0.465 N \Gamma$, corresponding to $\cos \tilde{\theta}_J \approx 0.5$, with an expected relative ratio of OAT to collective dephasing rates of 0.44, i.e., in the limit of significant dephasing. We consider 500 trajectories with time steps $dt \leq (N \Gamma)^{-1}/250$, which we reduce by a factor of 5 briefly after each time the drive is turned on or off, limiting the quantum jump probability at a given step to under 5\%. For the analytics, we consider a fixed, steady-state value $\cos \tilde{\theta}_J \approx 0.5$. 

We find excellent agreement, with the analytics ($-7.3$ dB) appearing to underestimate the exact results (minimum: $-7.8$ dB; steady-state: $-7.6$ dB). We attribute this disagreement to non-adiabatic and finite-size effects that result in slight $\hat J$ correlations even after the drive is turned off, illustrated by the small nonzero steady-state $\langle e \rangle \approx 6 \times 10^{-3}$ after the drive is off. If instead of considering $\mathcal{C}$, we instead calculate the squeezing in the $|{\uparrow}\rangle,|{\downarrow}\rangle$ basis directly, we find a steady-state squeezing of $-7.1$ dB, which differs from the analytic result by the same amount that is lost during the dynamics after the drive is turned off.

\end{document}


\title{Supplemental Material: Engineering One Axis Twisting via a Dissipative Berry Phase Using Strong Symmetries}	

\author{Jeremy T. Young}
\email[Corresponding author: ]{j.t.young@uva.nl}
\affiliation{Institute of Physics, University of Amsterdam, 1098 XH Amsterdam, the Netherlands}
\affiliation{JILA, University of Colorado and National Institute of Standards and Technology, and Department of Physics, University of Colorado, Boulder, Colorado 80309, USA}
\affiliation{Center for Theory of Quantum Matter, University of Colorado, Boulder, Colorado 80309, USA}

\author{Edwin Chaparro}
\affiliation{JILA, University of Colorado and National Institute of Standards and Technology, and Department of Physics, University of Colorado, Boulder, Colorado 80309, USA}
\affiliation{Center for Theory of Quantum Matter, University of Colorado, Boulder, Colorado 80309, USA}

\author{Asier Pi\~{n}eiro Orioli}
\affiliation{JILA, University of Colorado and National Institute of Standards and Technology, and Department of Physics, University of Colorado, Boulder, Colorado 80309, USA}
\affiliation{Center for Theory of Quantum Matter, University of Colorado, Boulder, Colorado 80309, USA}

\author{James K. Thompson}
\affiliation{JILA, University of Colorado and National Institute of Standards and Technology, and Department of Physics, University of Colorado, Boulder, Colorado 80309, USA}

\author{Ana Maria Rey}
\affiliation{JILA, University of Colorado and National Institute of Standards and Technology, and Department of Physics, University of Colorado, Boulder, Colorado 80309, USA}
\affiliation{Center for Theory of Quantum Matter, University of Colorado, Boulder, Colorado 80309, USA}

\date{\today}

\maketitle
\onecolumngrid

\renewcommand{\theequation}{S\arabic{equation}}
\renewcommand{\thesubsection}{S\arabic{subsection}}
\renewcommand{\thesubsubsection}{\Alph{subsubsection}}

\renewcommand{\bibnumfmt}[1]{[S#1]}
\renewcommand{\citenumfont}[1]{S#1} 

\pagenumbering{arabic}

\makeatletter
\renewcommand{\thefigure}{S\@arabic\c@figure}
\renewcommand \thetable{S\@arabic\c@table}

This supplemental material is organized as follows: In Sec.~\ref{sec:cavity}, we present the full atom-cavity model and adiabatically eliminate the cavity mode to derive the spin model of the main text. In Sec.~\ref{sec:MFT}, we investigate the mean-field theory behavior of the two-level system. In Sec.~\ref{sec:coherence}, we describe the formalism of weak and strong symmetry breaking in open quantum systems and discuss the preservation of the $\hat S$ coherences in our protocol from three perspectives.  
In Sec.~\ref{sec:weak}, we discuss the weak drive limit of our protocol, where the excited state can be adiabatically eliminated. In Sec.~\ref{sec:HP}, we implement the generalized Holstein-Primakoff approximation and derive effective Hamiltonian and dephasing dynamics beyond the adiabatic $\delta$ regime. In Sec.~\ref{sec:decoherence}, we optimize the squeezing in the presence of single-particle decoherence and compare this to other typical optical cavity squeezing protocols. 

\section{Cavity Model}
\label{sec:cavity}

Here, we discuss the behavior of the cavity photons, which have been adiabatically eliminated in the main text. To understand how their behavior is related to the spin model, we reconsider the full cavity-spin model
\begin{subequations}
\begin{equation}
    \partial_t \hat \rho = -i [\hat H, \hat \rho] + \kappa (\hat a \hat \rho \hat a^\dagger - \frac{1}{2} \{ \hat a^\dagger \hat a, \hat \rho \}),
\end{equation}
\begin{equation}
    \hat H = i \epsilon (\hat a^\dagger - \hat a) - \delta \hat N_e + g_c (\hat a \hat J^+ + \hat a^\dagger \hat J^-),
\end{equation}
\end{subequations}
where $\kappa$ represents the rate of photon leakage and the Tavis-Cummings term $g_c (\hat a \hat J^+ + \hat a^\dagger \hat J^-)$ describes the atom-photon interaction with single-photon Rabi frequency $2 g_c$. Here, we have assumed that the cavity is coherently driven via $\epsilon$, which we show leads to the drive in the spin model. Below, we also discuss how driving the spins directly modifies the analysis.

The mean-field dynamics for $\hat a, \hat J^-$ are given by
\begin{subequations}
\begin{equation}
    \partial_t \langle \hat a \rangle  =  \epsilon + i g_c \langle \hat J^-\rangle  - \frac{\kappa}{2} \langle \hat a\rangle,
\end{equation}
\begin{equation}
    \partial_t \langle \hat J^- \rangle = - i \delta \langle \hat J^- \rangle + 2i g_c \langle \hat a \rangle \langle \hat J_z \rangle,
\end{equation}
\end{subequations}
whose cavity steady-state solution is 
\begin{equation}
     \langle \hat a\rangle = \frac{\epsilon + i g_c \langle \hat J^- \rangle}{\kappa/2}.
\end{equation}
In the bad cavity limit $\kappa \gg \sqrt{N_J}g_c, \delta$, we can adiabatically eliminate the cavity by replacing it with its expectation value for the spin equations of motion, comparing to the corresponding mean-field spin dynamics of the main text
\begin{equation}
    \partial_t \langle \hat J^- \rangle = - i \delta \langle \hat J^- \rangle + \frac{4 i \epsilon g_c}{\kappa} \hat J_z - \frac{4 g_c^2}{\kappa} \langle \hat J_z \rangle \hat \langle J^- \rangle = - i \delta \langle \hat J^- \rangle + i \Omega \hat J_z - \Gamma \langle \hat J_z \rangle \langle \hat J^- \rangle, 
\end{equation}
which allows us to identify $\Omega = 4 \epsilon g_c/\kappa$ and $\Gamma = 4g_c^2/\kappa$. Hence we find the cavity steady state may also be expressed
\begin{equation}
\label{eq:aMF}
    \langle \hat a \rangle = i \frac{g_c}{\kappa/2}(\langle \hat J^-\rangle + i\Omega/\Gamma).
\end{equation}
This takes a value of 0 in the spin-polarized phase when $\delta = 0$, corresponding to the absence of photons. Because this holds for all values of $N_J$, this implies conservation of coherences in the steady state due to the environment's inability to distinguish the different values of $N_J$ for the spins.

Note that if the spins were driven directly, the constant term would be dropped in the above expression, and the cavity coherence would likewise be shifted to a constant, non-zero value $-2 g_c \Omega/(\kappa \Gamma)$, again independent of $N_J$. 

\section{Mean field theory}
\label{sec:MFT}

In this section, we determine the steady-state behavior of the two-level system using mean-field theory. As in the main text, we utilize a Schwinger boson approach, where $\hat d, \hat e$ denote the annihilation operators for $|{\downarrow}\rangle$ and $|e\rangle$ with corresponding mean-field values $d = e^{i \omega_Bt} \langle \hat d \rangle, e = e^{i \omega_Bt}  \langle \hat e \rangle$, with $\omega_B$ denoting the rotating frame associated with the non-trivial Berry phase. The resulting mean-field equations are
\begin{subequations}
    \begin{gather}
    \label{eq:MFTg}
        \partial_t d = - i \frac{\Omega}{2} e + i \omega_B d + \frac{\Gamma}{2} |e|^2 d = 0,\\
    \label{eq:MFTe}
        \partial_t e = - i \frac{\Omega}{2} d + i (\delta + \omega_B) e - \frac{\Gamma}{2} |d|^2 e = 0,
    \end{gather}
\end{subequations}
where we have assumed a drive in the $x$-direction as in $\hat H_\delta$ of the main text.

First, we identify $\omega_B$. Subtracting the $\Omega$ terms from each side and multiplying these two equations, we find
\begin{equation}
    \frac{\Omega^2}{4} = \left[ \frac{\Gamma}{2} |e|^2 + i \omega_B\right] \left[ \frac{\Gamma}{2} |d|^2 - i (\delta + \omega_B) \right].
\end{equation}
Noting that we have assumed $\Omega$ to be real without loss of generality, the two terms on the RHS have opposite complex phases, enforcing the condition
\begin{equation}
    \frac{|e|^2}{|d|^2} = \frac{\omega_B}{\delta + \omega_B},
\end{equation}
whose solution is readily solved
\begin{equation}
    \omega_B= \delta \frac{|e|^2}{|d|^2-|e|^2} = \frac{\delta}{2 \cos \theta_J} - \frac{\delta}{2},
\end{equation}
where we have used the fact that $|e|^2 = N \sin^2\frac{\theta_J}{2}, |d|^2 = N \cos^2 \frac{\theta_J}{2}$ on the Bloch sphere.

Next, we identify the steady-state values of $e, d$. We multiply Eq.~(\ref{eq:MFTg}) by $d^*$ and the complex conjugate of Eq.~(\ref{eq:MFTe}) by $e$ and subtract the two, finding
\begin{subequations}
    \begin{equation}
        i \Omega e d^* = i \delta |e|^2 + i \omega_B(|d|^2 + |e|^2) + \Gamma |d|^2 |e|^2,
    \end{equation}
    \begin{equation}
        \frac{\Omega}{2}  e^{-i \phi_J} \sin \theta_J=  \frac{\delta}{2} \sin \theta_J\tan \theta_J - i \frac{N_J \Gamma}{4} \sin^2 \theta_J,
    \end{equation}
\end{subequations}
where $e^* d = \langle \hat J^+ \rangle = \frac{N_J}{2} e^{i \phi_J} \sin \theta_J$.
Considering the real and imaginary parts separately, we find
\begin{subequations}
    \begin{equation}
        \Omega \cot \theta_J\cos \phi_J = \delta,
    \end{equation}
    \begin{equation}
        \frac{\Omega}{2} \sin \phi_J = \frac{N_J \Gamma}{4} \sin \theta_J,
    \end{equation}
\end{subequations}
from which $\theta_J, \phi_J$ may be identified. For the particular case of $\delta = 0$, we have $\sin \theta_J= 2 \Omega/N_J \Gamma = \Omega/\Omega_c$, $\phi_J = \pi/2$, as anticipated.

\section{Coherence preservation}
\label{sec:coherence}

In this section, we discuss the preservation of coherences. First, we discuss the formal definition of weak and strong symmetries and discuss the implications for the Liouvillian structure and the effect of this structure for spontaneous symmetry breaking. Second, we show this preservation under mean-field theory using Schwinger bosons. Third, we analytically show that for the observables of interest, all relevant coherences are preserved when the drive is turned off.

\subsection{Strong symmetry breaking}

To understand the distinction between weak and strong symmetries, it is helpful to reconsider the dynamics from the perspective of the full system-environment Hamiltonian, which takes the general form
\begin{equation}
    \hat H_{SE} = \hat H_S + \hat H_E + \hat V, \qquad \hat V = \sum_i \hat L_i \otimes \hat E_i,
\end{equation}
where $\hat H_S$ ($\hat H_E$) is the system (environment) Hamiltonian while $V$ is the interaction between the two composed of a product of system operators ($\hat L_i$) and environment operators ($\hat E_i$), and we have anticipated that the system operators in $\hat V$ will give rise to the Lindblad operators, which can always be expressed in a traceless, orthogonal basis.
Throughout, we only consider symmetries with separable unitary transformations, i.e., $\hat U = \hat U_S \otimes \hat U_E$, which commute with the full Hamiltonian $[\hat H_{SE}, \hat U] = 0$. In deriving the Liouvillian dynamics of the system, the environment is traced out. We can thus distinguish two relevant scenarios for symmetries of $H_{SE}$ that affect system: (i) The symmetry acts on both system and environment and (ii) The symmetry acts on only the system (i.e., $\hat U_E$ is the identity). 

In (i), information about the symmetry sectors and their conserved quantities is lost when the environment is traced out and $\hat U_E$ is no longer accessible, so the conservation law can at most be satisfied in the system after ensemble/trajectory averaging. 
In (ii), no information about the symmetry is lost when the environment is traced out since $\hat U_E$ is the identity, so the corresponding conserved quantity must be conserved on an individual trajectory level within a given symmetry sector of the system. From this, we can understand (i) as corresponding to weak symmetries of the Liouvillian, while (ii) corresponds to strong symmetries, although we remark that the Hamiltonian framework here can be extended beyond Liouvillian dynamics \cite{Manzano2018}. Recently, connections and equivalencies have further been identified between the distinction of weak/strong symmetries and symmetries of Hamiltonians in the presence of disorder \cite{Ma2023a,Ma2023b}. Specifically, the concept of an average symmetry emerges in disordered systems when the symmetry is violated for any individual realization but is restored upon ensemble averaging of many realizations. The disorder averaging is thus analogous to the trajectory averaging of a weak symmetry and vice-versa.

With this in mind, we can see the implications of these two scenarios on the Liouvillian 
\begin{equation}
    \partial_t \hat \rho = \mathcal{L}(\hat \rho), \qquad
    \mathcal{L}(\cdot) \equiv -i[\hat H_S,(\cdot)] + \sum_i \hat L_i(\cdot) \hat L_i^\dagger - \frac{1}{2} \{\hat L_i^\dagger \hat L_i, (\cdot)\}
\end{equation} 
where $\hat L_i$ are the Lindblad jump operators, and $\hat A \equiv \hat U_S$ is the corresponding symmetry operator of the system. Since $\mathcal{L}$ is a superoperator, we denote its action on a generic operator via inserting the operator anywhere $(\cdot)$ occurs. 
Thus in the case of (i), a weak symmetry, since $\hat U_S \hat V \hat U_S^\dagger \neq \hat V$ (the action of $\hat U_E$ is necessary for commutation), we have $[\hat L_i, \hat A] \neq 0$. Nevertheless, although we do not prove it here,  the Lindblad operators can always be expressed in a new basis such that the action of the unitary is only to apply a phase shift to the $\hat L_i$, $\hat A \hat L_i \hat A^\dagger = e^{i \phi} \hat L_i$,. 
Since unlike in $\hat H_{SE}$, the Lindblad operators appear in pairs,  the combined transformation of both can leave the Liouvillian itself invariant given that the phase from $\hat L_i$ and $\hat L_i^\dagger$ cancel. 
We thus define the symmetry superoperator $\mathcal{A}(\cdot) = \hat A (\cdot) \hat A^\dagger$ in order to capture the fact that both jump operators must transform under $\hat A$, where the symmetry of $\mathcal{L}$ is now captured by its commutation with the symmetry superoperator $[\mathcal{L},\mathcal{A}] \equiv \mathcal{L}(\mathcal{A}(\cdot)) - \mathcal{A}(\mathcal{L}(\cdot)) = 0$. The parentheses in this notation indicate the order of the action of superoperators, e.g., $\mathcal{L}(\mathcal{A}(\cdot))$ indicates that we apply $\mathcal{A}$ then $\mathcal{L}$ to the generic operator in $(\cdot)$. If the operators are vectorized, the superoperators can be expressed as matrices, in which case this commutation relation reduces to the usual definition in terms of matrices.

Like for a Hamiltonian, $\mathcal{L}$ is block diagonal in the symmetry sectors of $\mathcal{A}$ for a weak symmetry after vectorizing $\hat \rho$, 
\begin{equation}
    \mathcal{L} = \left(
    \begin{array}{ccc}
    \mathcal{L}_1 &  & 0 \\
     & \ddots &  \\
    0 &  & \mathcal{L}_{n} 
    \end{array}
    \right),
\end{equation}
where $n$ is the number of symmetry sectors.
There is always a single block with eigenvalue 1 containing all operators with non-vanishing trace, denoted $\mathcal{L}_1$, while the remaining sectors are all composed of traceless operators. For example, for two symmetry sectors of $\hat A$ $|a \rangle, |b \rangle$, e.g., parity, there are two symmetry sectors of $\mathcal{A}$  (see Ref.~\cite{Buca2012} for a generic number of symmetry sectors). The first, corresponding to $\mathcal{L}_1$, includes operators of the form $|a \rangle \langle a|, |b \rangle \langle b|$, and have eigenvalue 1 (the two phases from the symmetry operator on each side cancel), and all operators (and thus density matrices) with nonzero trace are in this sector since the other sectors involve bras and kets of different symmetry sectors. The second, corresponding to $\mathcal{L}_2$, includes operators of the form $|a\rangle \langle b|, |b \rangle \langle a|$, all of which are traceless and have conjugate eigenvalues other than 1 (the two phases do not cancel). In principle $|a\rangle \langle b|, |b \rangle \langle a|$ can define two conjugate symmetry sectors, but since they are equivalent we may combine them into one. 

While the generic conditions on the uniqueness of the steady state, particularly in the thermodynamic limit, are complicated in general (see, e.g., Ref.~\cite{Zhang2024} for an overview), for most typical open quantum systems with only a weak symmetry we expect that $\mathcal{L}_1$ will always have a 0 eigenoperator corresponding to a steady state at any system size and for any system parameters. Moreover, we expect that in the thermodynamic limit, when spontaneous symmetry breaking of the weak symmetry occurs, at least one of the $\mathcal{L}_{i>1}$ realizes a traceless 0 eigenoperator, as in the example investigated in Ref.~\cite{Lieu2020}.
This is analogous to the transverse-field Ising model, where the ground state is symmetric (eigenvalue 1) for a finite system, while the antisymmetric (eigenvalue $-1$) ground state becomes degenerate only in the thermodynamic limit. 

In the case of (ii), a strong symmetry, we have $[\hat L_i, \hat A] = 0$. This means either of the pairs of jump operators in the master equation can be transformed while leaving the master equation invariant. Therefore, in contrast to the weak symmetry we can instead define \textit{two} symmetry superoperators $\mathcal{A}_l(\cdot) \equiv \hat A(\cdot)$, $\mathcal{A}_r (\cdot) \equiv (\cdot) \hat A^\dagger$, both of which satisfy $[\mathcal{L}, \mathcal{A}_{l/r}] = 0$ and correspond to transformations on operators on either the left or right hand side of the Liouvillian.

If there are $n$ symmetry sectors of $\hat A$, then these two symmetry superoperators block diagonalize $\mathcal{L}$ into $n^2$ blocks. Without loss of generality, we can consider two sectors $|a_1\rangle, |a_2\rangle$ denoting all states with eigenvalues $a_1, a_2$ (e.g., eigenvalues of $e^{i \hat N_J/N}$). We may thus write 
\begin{equation}
    \mathcal{L} = \left(
    \begin{array}{cccc}
    \mathcal{L}_{11} &  &  & 0 \\
     & \mathcal{L}_{22} &  &  \\
     &  &\mathcal{L}_{12} &   \\
    0 &  &  & \mathcal{L}_{21}
    \end{array}
    \right),
\end{equation}
where the blocks correspond to the spaces $|a_1\rangle \langle a_1|,|a_2\rangle \langle a_2|,|a_1\rangle \langle a_2|$, and $|a_2\rangle \langle a_1|$, in order. In this case, both $\mathcal{L}_{11}$ and $\mathcal{L}_{22}$ have non-vanishing trace, and both will possess eigenoperators with eigenvalue 0 for most typical open quantum systems which exhibit steady states. This is a manifestation of the conservation law for any individual trajectory: if the system is initialized in one symmetry sector, it must remain there. This indicates that there are generally multiple steady states even for typical finite systems, one for each sector, as recently observed experimentally \cite{Li2023}. The eventual projection into a given symmetry sector on an individual trajectory level has been termed dissipative freezing \cite{SanchezMunoz2019} and can be understood as arising due to a weak measurement of the symmetry. We remark that the behavior of trajectories in the unbroken symmetry phase, such as for dissipative freezing, depends on the precise unraveling. For example, while a weak measurement and a dephasing term can have equivalent Liouvillian dynamics, if the unraveling for the dephasing is achieved through white noise on a Hamiltonian term, there will be no such projection. 

Going beyond the unbroken symmetry phase, we expect spontaneous symmetry breaking of the strong symmetry to correspond to $\mathcal{L}_{12},\mathcal{L}_{21}$ (which are equivalent) obtaining an eigenoperator with 0 eigenvalue in the thermodynamic limit, as in Ref.~\cite{Lieu2020}. Since these correspond to coherences between $|a_1\rangle$ and $|a_2\rangle$, symmetry breaking indicates the preservation of coherences in the steady state, as captured by its association with passive error correction \cite{Lieu2020,Liu2023}, which has likewise enabled the extension of symmetry protected topological phases in open quantum systems \cite{deGroot2022} and the investigation of the effect of the type of symmetry on criticality in these systems \cite{Minganti2023}. This feature of the strong symmetry thus corresponds to the emergence of a decoherence free subspace or a more general noiseless subsystem \cite{Lidar1998, Knill2000, Albert2014, Lieu2020} and enables the near-preservation and generation of the different relative Berry phases throughout the protocol.

\subsection{Mean field theory}

The additional coherences of interest are $\hat u^\dagger \hat g, \hat u^\dagger \hat e$, where $\hat u$ denotes the collective annihilation operator for the $|{\uparrow}\rangle$ state with corresponding mean-field value $u = \langle \hat u \rangle$. Since $\hat u$ commutes with the Hamiltonian and jump operator, we find $\partial_t u = 0$. Hence, we find
\begin{equation}
    \partial_t \langle \hat u^\dagger \hat d \rangle_{\text{MF}} = u^* \partial_t d, \qquad \partial_t \langle \hat u^\dagger \hat e \rangle_{\text{MF}} =  u^* \partial_t e,
\end{equation}
indicating that the evolution of the coherences within $\hat S$ are determined purely by those of $\hat J$ at the mean-field level. Moreover, if we look at the magnitude of these two coherences $\mathcal{U} \equiv (\hat u^\dagger \hat d, \hat u^\dagger \hat e)$, we find
\begin{equation}
    \partial_t |\langle \mathcal{U} \rangle_{\text{MF}} |^2 = |u|^2 \partial_t (|d|^2+|e|^2) = |u|^2 \partial_t N = 0,
\end{equation}
indicating these coherences are preserved under mean-field theory and only change their phase information. As a result, their evolution is otherwise determined entirely by $d, e$. In the spin-polarized phase, they relax to a steady-state value while maintaining the original coherence of the initial state up to deterministic phase factors. In the mixed phase, the modified Rabi oscillations impart themselves on these coherences as well. This mirrors the two-level model, where these oscillations are a consequence of the conservation of $\hat J^2$, which implies mean-field conservation of $\langle \hat J_x \rangle_{\text{MF}}^2 + \langle \hat J_y \rangle_{\text{MF}}^2 + \langle \hat J_z \rangle_{\text{MF}}^2$, preventing mean-field theory from directly capturing the loss of coherence in the steady state. Thus these oscillations in the three-level coherences imply a similar loss of coherence in the steady state.

\subsection{Turning off the drive}

Next, we discuss the preservation of coherence when the drive is turned off. While we expect that the results here qualitatively hold for any quenches of $\Omega$ within the spin-polarized phase for $\delta \to 0$, the case of no drive can be solved exactly for any initial density matrix. 

First, we note that only coherences between states with an equal number of $|e\rangle$ excitations can be preserved in the $|{\uparrow}\rangle, |{\downarrow}\rangle$ manifold because the recycling term in the Lindbladian always removes excitations from both the ket and the bra simultaneously. Physically, if two states emit different numbers of photons, then they are perfectly distinguished, preventing the conservation of any coherence. Writing our states in the permutationally symmetric basis $|n_{\downarrow}, n_{\uparrow}, n_e \rangle$, where $n_\mu$ denotes the number of atoms in $|\mu\rangle$, we find
\begin{subequations}
\begin{equation}
    \Gamma^{-1} \partial_t \langle \mathcal{C}_{n_e} \rangle =  w_{n_e+1} \langle \mathcal{C}_{n_e} \rangle - W_{n_e} \langle \mathcal{C}_{n_e} \rangle, \qquad \mathcal{C}_{n_e} \equiv |n_{\downarrow}, n_{\uparrow}, n_e \rangle \langle n_{\downarrow}', n_{\uparrow}', n_e |
\end{equation}
\begin{equation}
    w_{n_e} \equiv \sqrt{(n_{\downarrow}+1) n_e  (n_{\downarrow}' + 1) n_e} , \qquad W_{n_e} \equiv \frac{(n_{\downarrow}+1) n_e + (n_{\downarrow}' + 1) n_e}{2}
\end{equation}
\end{subequations}
where we have dropped $n_{\updownarrow}$ labels in $\mathcal{C}_{n_e}$ for simplicity. Thus a coherence with $n_e$ excitations decay at rate $W_{n_e}$ while gaining coherence from the corresponding $n_e+1$ excitation coherence at a rate $w_{n_e+1}$. We remark that $w_{n_e}$ and $W_{n_e}$ are the geometric and arithmetic means of the superradiant decay rates (in units of $\Gamma$) of the bra and ket with $n_e$ excitations.

The full rate equation is expressed
\begin{equation}
    \Gamma^{-1} \partial_t \vec{\mathcal{C}} = 
    \left(
    \begin{array}{cccc}
        \ddots & \ddots & \ddots & \vdots \\
        \ddots & -W_{2} & \ddots & 0 \\
        \ddots & w_2 & -W_1 & 0 \\
        \cdots & 0 & w_1 & 0
    \end{array}
    \right) \vec{\mathcal{C}}, \qquad \vec{\mathcal{C}} \equiv ( \cdots, \mathcal{C}_2, \mathcal{C}_1, \mathcal{C}_0)^T.
\end{equation}
Since we are interested in the steady state, we identify the left ($\vec{u}_i$) and right ($\vec{v}_i$) eigenvectors with 0 eigenvalue $W_0 = 0$
\begin{subequations}
\begin{equation}
    \vec{v}_0 = (0, \cdots, 0, 1),
\end{equation}
\begin{equation}
    \vec{u}_0 = ( \cdots, \beta_2 \beta_1, \beta_1, 1), \qquad \beta_{n_e} \equiv \frac{w_{n_e}}{W_{n_e}},
\end{equation}
\end{subequations}
which satisfy the orthogonality condition $\vec{u}_i \cdot \vec{v}_j = \delta_{ij}$. We identify the final coherence $\mathcal{C}_0$ by noting that if $\vec{\mathcal{C}}(t) = \sum_{n_e} a_{n_e} \vec{v}_{n_e} e^{-W_{n_e} t}$, then the orthogonality condition implies $\mathcal{C}_0 (t \to \infty) = a_0 = \vec{u}_0 \cdot \vec{\mathcal{C}}(0)$. From this, we see that the $\beta_{n_e} \leq 1$ thus represent the fraction of the $n_e$ coherence which is converted into $n_e - 1$ coherence. Given the arithmetic and geometric mean form of $w$ and $W$, we see that $\beta$ represents a measurement of the equivalence of the superradiant decay rates, i.e., their distinguishability. Expanding $\beta_{n_e}$ in powers of $n_{\downarrow} - n_{\downarrow}' = N_J - N_J'$, noting $n_{\downarrow} = N_J - n_e$,
\begin{subequations}
\begin{equation}
    \beta_{n_e} \approx 1 - \frac{(N_J - N_J')^2}{8 (N_J - n_e+1)^2},
\end{equation}
\begin{equation}
    \prod_{n_e' = 1}^{n_e} \beta_{n_e'} \approx 1 - \sum_{n_e' = 1}^{n_e}\frac{(N_J - N_J')^2}{8 (N_J - n_e+1)^2} \approx 1 -\frac{(N_J-N_J')^2}{8} \int_0^{n_e} \frac{dx}{(N_J - x)^2} = 1 - \frac{n_e}{8 N_J (N_J - n_e)} (N_J-N_J')^2.
\end{equation}
\end{subequations}
From this, we see that if $(N_J - N_J)' \ll \mathcal{O}(\sqrt{N})$, the fraction of coherences with $n_e$ excitations that survives converges to 1 for large $N$. Since spin squeezing in the $|{\uparrow}\rangle,|{\downarrow}\rangle$ manifold only depends on coherences between $|N_J - N_J'| \leq 2$, we expect nearly perfect transfer of the spin squeezing after the drive is turned off.

\section{Weak drive limit}
\label{sec:weak}

In this section, we derive the effective Hamiltonian and jump operator in the limit of weak drive. In this limit, the excited state $|e\rangle$ can be adiabatically eliminated, leaving only the effective dynamics of the ground-state manifold. To do so, we utilize a perturbative approach applicable to dissipative systems \cite{Reiter2012}. Furthermore, we will consider the role of collective interactions on the model which arise due to a detuning $\Delta$ of the cavity from the atomic transition. The corresponding Hamiltonian is
\begin{equation}
    \hat H = \Omega \hat J_x - \delta \hat N_e + \chi \hat J^\dagger \hat J, \qquad \hat l = \hat J^-, \qquad \chi = \frac{g_c^2}{\Delta^2+\kappa^2/4} \Delta, \qquad \Gamma_{\Delta} = \frac{g_c^2}{\Delta^2+\kappa^2/4} \kappa, 
\end{equation}
where $2 g_c$ is the single-photon Rabi frequency of the $|{\downarrow}\rangle \to |e\rangle$ transition and $\kappa$ is the decay rate of the cavity. Here, we have introduced $\Gamma_{\Delta}$ to distinguish it from the superradiant decay rate $\Gamma = \Gamma_{\Delta = 0}$ of the main text, where $\Delta = 0$.

To lowest non-trivial (second-) order, this results in the general expressions for effective Hamiltonian and jump operator 
\begin{subequations}
\begin{equation}
    \hat H_\updownarrow \approx - \frac{1}{2} \hat V_- [\hat H_{{{\text{NH}}}}^{-1} + (\hat H_{{{\text{NH}}}}^\dagger)^{-1}]\hat V_+, \qquad \hat l_\updownarrow = \hat l \hat H_{{{\text{NH}}}}^{-1} \hat V_+,
\end{equation}
\begin{equation}
    \hat H_{\text{NH}} \equiv \hat H_e - \frac{i}{2} \hat l^\dagger \hat l,
\end{equation}
\end{subequations}
where $\hat V^+ = \Omega/2 \hat J^+$ and its conjugate are the perturbative terms and $\hat H_e$ encodes the excited-state energies relative to the coupled ground state. Here, we see that the non-Hermitian $\hat{H}_{\text{NH}}^{-1} + H.c.$ replaces the usual inverse energy difference in conventional Hamiltonian perturbation theory while incorporating the non-zero linewidths of the bare Hamiltonian's eigenstates.

For the three-level model and fixed $N_J$, the non-Hermitian Hamiltonian takes the form
\begin{subequations}
\begin{equation}
    \hat H_{{{\text{NH}}}} = -\delta \left(\frac{N_J}{2} + \hat J_z\right) - \left(\chi + i \frac{\Gamma_\Delta}{2} \right) \hat J^+ \hat J^- = -\delta \left(\frac{N_J}{2} + \hat J_z\right) - \left(\chi + i \frac{\Gamma_\Delta}{2} \right) \left[ \hat J^2 - \hat J_z^2 + \hat J_z \right].
\end{equation}
\end{subequations}
Since we are interested in the weak drive limit where all atoms are in the ground-state manifold, we restrict our focus to the state $|J = N_J/2, m_J = -N_J /2\rangle$, where $J$ and $m_J$ are the total and azimuthal quantum numbers of $\hat{\mathbf{J}}$. Thus we find the resulting (complex) Stark shift in this limit to be
\begin{equation}
    \hat J^- \hat H_{{\text{NH}}}^{-1} \hat J^+ |N_J/2,-N_J/2 \rangle = -\frac{N_J}{\delta +(\chi + i \frac{\Gamma_\Delta}{2}) N_J} |N_J/2,-N_J/2 \rangle,
\end{equation}
which is non-linear in $N_J$ when $\delta \neq 0$. Both $\chi, \Gamma_\Delta$ give rise to nonlinear fluctuations (in $N_J$) to the Stark shifts in a similar way, although $\Gamma_\Delta$ does so via the linewidths.

In the weak drive limit, we restore fluctuations in $N_J$ according to $\hat{N}_J \equiv N/2 - \hat{S}_{\updownarrow,z}$, where $\updownarrow$ denotes the $|{\uparrow},{\downarrow}\rangle$ manifold, resulting in the following effective Hamiltonian and jump operator
\begin{equation}
\begin{aligned}
    \hat H_{\updownarrow} &~= \frac{\Omega^2}{4} \frac{[\delta+\chi  (N/2-\hat{S}_{\updownarrow,z})] (N/2-\hat{S}_{\updownarrow,z})}{[\delta+\chi  (N/2-\hat{S}_{\updownarrow,z})]^2+\frac{\Gamma_\Delta^2}{4} (N/2-\hat{S}_{\updownarrow,z})^2}, \\
    \hat l_{\updownarrow} &~= -\frac{\Omega}{2}\frac{N/2-\hat{S}_{\updownarrow,z}}{\delta +(\chi + i \frac{\Gamma_\Delta}{2}) (N/2-\hat{S}_{\updownarrow,z})}
    \\ &~ = -\frac{\Omega}{2} \left( -\frac{\delta \hat{S}_{\updownarrow,z}}{[\delta + (\chi + i \Gamma_\Delta/2) N/2][ \delta + (\chi + i \Gamma_\Delta/2) (N/2 - S_{\updownarrow,z})]}  + \frac{N/2}{\delta + (\chi + i \Gamma_\Delta/2)N/2}\right).
\end{aligned}
\end{equation}
Here, we note that there is a constant term in the jump operator, which can give rise to unitary dynamics in combination with the operator part of the jump operator. For a jump operator $C + \hat O$, this corresponds to a Hamiltonian term $\Gamma_\Delta (C \hat O^\dagger - C^* \hat O)/2i$. Absorbing the effect of the constant term into the Hamiltonian, we find
\begin{subequations}
\begin{align}
    \hat H_{\updownarrow} &~ = \frac{\Omega^2}{4} \frac{[\delta+\chi  (N/2-\hat{S}_{\updownarrow,z})] (N/2-\hat{S}_{\updownarrow,z})}{[\delta+\chi  (N/2-\hat{S}_{\updownarrow,z})]^2+\frac{\Gamma_\Delta^2}{4} (N/2-\hat{S}_{\updownarrow,z})^2}\left(1 - \frac{N \Gamma_\Delta^2 \delta \hat S_{\updownarrow,z} /4 }{[\delta+ \chi (N/2+\hat S_{\updownarrow,z})] [ (\delta +N \chi/2)^2 + (N \Gamma_\Delta/4)^2] }\right),\\
    \hat l_{\updownarrow} &~ = \frac{\Omega}{2} \left( \frac{\delta \hat{S}_{\updownarrow,z}}{[\delta + (\chi + i \Gamma_\Delta/2) N/2][ \delta + (\chi + i \Gamma_\Delta/2) (N/2 - S_{\updownarrow,z})]}\right).
\end{align}
\end{subequations}

Finally, we can expand both in powers of $\hat S_{\updownarrow,z}$ when assuming an equal superposition of the ground states ($\langle \hat S_{\updownarrow,z} \rangle = 0$). Expanding to second order in the master equation (second order in the Hamiltonian and first order in the jump operator), we find
\begin{subequations}
    \begin{align}
        \hat H_{\updownarrow} &~ \approx \frac{N \Omega^2 (\delta + N \chi/2) }{8| \delta + N \chi/2 + i N \Gamma_\Delta/4|^2} - \frac{ \delta \Omega^2 }{4 | \delta + N \chi/2 + i N \Gamma_\Delta/4|^2} \hat S_{\updownarrow,z} - \frac{ \delta \Omega^2 (N \Gamma_\Delta^2/16 +  \delta \chi/2 + N \chi^2/4) }{2 |\delta + N \chi/2 + i N \Gamma_\Delta /4|^4} \hat S_{\updownarrow,z}^2 \\
        \hat l_{\updownarrow} &~ \approx  \frac{\delta \Omega }{2[ \delta + (\chi + i \Gamma_\Delta/2) N/2]^2}\hat S_{\updownarrow,z},
    \end{align}
\end{subequations}
whose ratio of OAT to collective dephasing rates is
\begin{equation}
    \frac{2 (\delta \chi/2 + N \chi^2/4 + N\Gamma_\Delta^2/16)}{\delta \Gamma_\Delta},
\end{equation}
which approaches $\chi/\Gamma_\Delta = \Delta/\kappa$ in the limit of large $\delta$ and $N g_c^2 /(4 \delta \kappa)$ in the limit of near-resonant drive. In the limit $\chi = 0, \delta \to 0$, we find agreement with the Berry phase derivations of the previous section.

\section{Holstein-Primakoff}
\label{sec:HP}

In this section, we derive the Holstein-Primakoff Hamiltonian and jump operator. To do so, we utilize the transformation 
\begin{equation}
    \left( 
    \begin{array}{c}
    \hat c^\dagger \\
    \hat s^\dagger \\
    \hat \jmath^\dagger 
    \end{array} 
    \right)
    = 
    \left(
    \begin{array}{ccc}
         \sqrt{f_J} \cos \frac{\tilde{\theta}_J}{2} & \sqrt{f_J} e^{-i \tilde{\phi}_J} \sin \frac{\tilde{\theta}_J}{2} & \sqrt{f_{\uparrow}} \\
         -\sqrt{f_{\uparrow}} \cos \frac{\tilde{\theta}_J}{2} & -\sqrt{f_{\uparrow}} e^{- i \tilde{\phi}_J} \sin \frac{\tilde{\theta}_J}{2} & \sqrt{f_J}\\
         \sin \frac{\tilde{\theta}_J}{2} & -e^{i \tilde{\phi}_J} \cos \frac{\tilde{\theta}_J}{2} & 0
    \end{array} \right)
    \left(
    \begin{array}{c}
        \hat d_{\tilde{\omega}_B}^\dagger \\
        \hat{e}_{\tilde{\omega}_B}^\dagger \\
        \hat u^\dagger     
    \end{array}
    \right),
\end{equation}
where $\hat c$ captures the mean-field steady state, $\hat \jmath$ the quantum fluctuations of the $\hat J$ Bloch sphere, and $\hat s$ the fluctuations of the $\hat S$ Bloch sphere. Here, the $\tilde{\omega}_B$ subscripts denote that we are in the rotating frame $\hat e_{\tilde{\omega}_B} = e^{i \tilde{\omega}_B t} \hat e, ~\hat d_{\tilde{\omega}_B} = e^{i \tilde{\omega}_B t} \hat d$, which introduces an additional Hamiltonian term $-\tilde{\omega}_B(\hat{e}^\dagger_{\tilde{\omega}_B} \hat{e}_{\tilde{\omega}_B} + \hat{d}^\dagger_{\tilde{\omega}_B} \hat{d}_{\tilde{\omega}_B})$. We also assumed without loss of generality that $\varphi = 0$ on the $\hat S$ Bloch sphere.

Expressing the Hamiltonian and jump operator in terms of $\hat c, \hat s, \hat \jmath$, we implement a generalized Holstein-Primakoff transformation according to the substitution $\hat c \to \sqrt{N - \hat{s}^\dagger \hat s - \hat{\jmath}^\dagger \hat \jmath}$, which we truncate at quadratic order in the Hamiltonian and first order in the jump operator. 
It is convenient to relate the quadratures of $\hat s, \hat \jmath$ to spin operators
\begin{subequations}
   \begin{gather}
        \sum_i  \frac{|0_i \rangle \langle 0_i| - |1_i \rangle\langle 1_i|}{2} \equiv
       \hat S_z \approx  \sqrt{\frac{N}{2}} \hat x_s, \qquad
        \sum_i \frac{|0_i \rangle \langle 1_i| - |1_i \rangle \langle 0_i|}{2i} 
       \equiv \hat S_y \approx -\sqrt{\frac{N}{2}} \hat p_s, \\
       \hat J_y \cos \tilde{\phi}_J - \hat J_x \sin \tilde{\phi}_J \approx \sqrt{\frac{f_J N}{2}} \hat p_j, \qquad
       (\hat J_x \cos \tilde{\phi}_J + \hat J_y \sin \tilde{\phi}_J ) \cos \tilde{\theta}_J + \hat J_z \sin \tilde{\theta}_J \approx -\sqrt{\frac{f_J N}{2}} \hat x_j
   \end{gather}
\end{subequations}
where we have chosen $f_J = f_{\uparrow} = 1/2$ so the steady state points in the $x$-direction on the $\hat S$ Bloch sphere. Here we see that the quadratures of the bosons directly correspond to fluctuations perpendicular to the Bloch vector on the $\hat S, \hat J$ Bloch spheres.
We further note that because nearly all population of the $J$ states is in the $|1\rangle$ state, $\hat S_z \approx (\hat N_{\uparrow} - \hat N_J)/2$, with both proportional to $\hat x_s$ at linear order in the Holstein-Primakoff expansion. Since the strong symmetry implies conservation of $\hat N_{\uparrow} - \hat N_{J}$, $\hat x_s$ will commute with both Hamiltonian and jump operator at the order we consider. 

Here, we note that as a result of this transformation, the jump operator takes the form $C + \hat O$ for some constant $C$ and operator $\hat O$ like in the weak drive limit. We can remove the constant term from the jump operator by shifting its effect to the Hamiltonian via an additional term $\Gamma (C \hat{O}^\dagger - C^* \hat O)/2i$.
In the new basis following the Holstein-Primakoff approximation, we identify the effective quadratic Hamiltonian
\begin{subequations}
\begin{multline}
    \hat H_{\text{HP}} \approx -\frac{\delta}{\cos \tilde{\theta}_J} \hat \jmath^\dagger \hat \jmath + \left[\sqrt{\frac{f_J N}{2}} \hat p_j - \sqrt{f_{\uparrow}} \frac{\hat s^\dagger \hat \jmath - \hat \jmath^\dagger \hat s}{2i} \right] \left( \frac{f_J N \Gamma}{2} \sin \tilde{\theta}_J - \Omega \sin \tilde{\phi}_J \right) +\\
    \left[\frac{f_J }{2}(N-\hat{s}^\dagger \hat s - \hat{\jmath}^\dagger \hat \jmath) - \sqrt{\frac{f_J f_{\uparrow} N}{2}} \hat x_s  + \frac{f_{\uparrow}}{2} \hat s^\dagger \hat s -\frac{1}{2} \hat \jmath^\dagger \hat \jmath \right] \sin \tilde{\theta}_J  \left( \Omega \cos \tilde{\phi}_J - \delta \tan \tilde{\theta}_J \right) + \\
    \left[-\sqrt{\frac{ f_J N}{2}}  \hat x_j  + \sqrt{f_{\uparrow}} \frac{\hat s^\dagger \hat \jmath + \hat \jmath^\dagger \hat s}{2}\right] \cos \tilde{\theta}_J  \left( \Omega \cos \tilde{\phi}_J - \delta \tan \tilde{\theta}_J \right),
\end{multline}
\begin{equation}
    \hat H_{\text{HP}} \approx -\frac{\delta}{\cos \tilde{\theta}_J} \hat \jmath^\dagger \hat \jmath,
\end{equation}
\end{subequations}
where most terms go to zero according to the mean field equations.
To linear order (quadratic in the master equation), the jump operator is
\begin{equation}
    \hat l_{\text{HP}} \approx -e^{- i \tilde{\phi}_J} \sqrt{\frac{f_J N}{2}}\left(\hat x_j \cos \tilde{\theta}_J + i \hat p_j + \sqrt{f_{\uparrow}} \hat x_s \sin \tilde{\theta}_J \right).
\end{equation}
As anticipated, the Hamiltonian and jump operator both commute with $\hat x_s$ as a consequence of the strong symmetry.

Next, we will derive the effective Hamiltonian and jump operators for only the $\hat s$ boson by adiabatically eliminating $\hat \jmath$. The equation of motion for $ \hat \jmath $ is
\begin{equation}
    \partial_t \hat \jmath \approx \left(i \delta \sec \tilde{\theta}_J - \frac{f_J N \Gamma}{2} \cos \tilde{\theta}_J \right) \hat \jmath  - \frac{f_J N \Gamma \sin \tilde{\theta}_J}{2}  \sqrt{\frac{f_{\uparrow}}{2}} \hat x_s.
\end{equation}
Here, we note that the above equation implies that $\hat \jmath$ will relax quickly to its steady state value, justifying the adiabatic elimination of it. By setting the evolution to 0 and solving for $\hat \jmath$, we make the substitution
\begin{equation}
    \hat \jmath \to \frac{f_J N \Gamma \sin \tilde{\theta}_J}{2 i \delta \sec \tilde{\theta}_J - f_J N \Gamma \cos \tilde{\theta}_J} \sqrt{\frac{f_{\uparrow}}{2}} \hat x_s,
\end{equation}
which results in the new Hamiltonian and jump operator
\begin{subequations}
    \begin{equation}
        \hat H_{\text{HP+AE}} \approx -\frac{\delta}{2 \cos \tilde{\theta}_J} \frac{ f_\uparrow f_J^2 N^2 \Gamma^2 \sin^2 \tilde{\theta}_J}{f_J^2 N^2 \Gamma^2 \cos^2 \tilde{\theta}_J + 4 \delta^2 \sec^2 \tilde{\theta}_J} \hat{x}_s^2,
    \end{equation}
    \begin{equation}
        \hat{l}_{\text{HP+AE}} \approx 2 e^{-i \tilde{\phi}_J} \sqrt{\frac{f_J f_{\uparrow} N}{2}} \frac{\delta \tan \tilde{\theta}_J (i f_J N \Gamma - 2 \delta \sec \tilde{\theta}_J) }{f_J^2 N^2 \Gamma^2 \cos^2 \tilde{\theta}_J + 4 \delta^2 \sec^2 \tilde{\theta}_J} \hat x_s.
    \end{equation}
\end{subequations}
Here, the Hamiltonian gives rise to shearing dynamics in the $\hat s$ boson in the $\hat p_s$ direction while the jump operator leads to dephasing. 

With this expression, we can use the associated timescales to self-consistently ensure that the adiabatic elimination is justified. Since the relaxation of $\hat \jmath$ is set by $f_J N \Gamma \cos \tilde{\theta}_J$, the squeezing dynamics should occur on timescales longer than this. Assuming we keep $\delta/N \Gamma$ fixed while varying $N$, then the Hamiltonian dynamics of $\hat H_{\text{AE}}$ scales like $N \Gamma$ while the OAT rate will have no $N$-dependence. Since the timescale of optimal squeezing scales slower than $N^{-2/3}$ (see Sec.~\ref{sec:decoherence}), the squeezing dynamics will be slower than the relaxational timescale of $\hat \jmath$, which scales like $N^{-1}$, for sufficiently large $N$. The precise condition for the validity of adiabatic elimination for any finite $N$ will depend on the choices of $\cos \tilde{\theta}_J$ and $\delta$ due to critical slowing down and higher-order effects beyond what we have considered here.

The relative rates of shearing to dephasing is given by
\begin{equation}
     \frac{ f_J N \Gamma \cos \tilde{\theta}_J}{ 4 |\delta|} \frac{f_J^2 N^2 \Gamma^2 \cos^2 \tilde{\theta}_J + 4 \delta^2 \sec^2 \tilde{\theta}_J^2}{f_J^2 N^2 \Gamma^2+4 \delta^2 \sec^2 \tilde{\theta}_J }, 
\end{equation}
which is maximized in the limit $f_J N \Gamma \cos^2 \tilde{\theta}_J \gg 2\delta$. In this limit, the effective Hamiltonian and jump operator can be simplified to
\begin{subequations}
    \begin{equation}
        \hat H_{\text{HP+AE}}\approx - \delta f_{\uparrow} \frac{\sin^2 \tilde{\theta}_J}{2 \cos^3 \tilde{\theta}_J} \hat x_s^2,
    \end{equation}
    \begin{equation}
        \hat l_{\text{HP+AE}} \approx 2 i \delta e^{- i \tilde{\phi}_J} \sqrt{\frac{f_J f_{\uparrow} N}{2}} \frac{\sin \tilde{\theta}_J}{ f_J N \Gamma \cos^3 \tilde{\theta}_J} \hat x_s.
    \end{equation}
\end{subequations}
The effective $\hat S$ dynamics in both cases can be obtained via the mapping $\hat x_s \to \sqrt{\frac{2}{N}} \hat S_z$ as in the main text, and we find the OAT in the small $\delta$ limit is consistent with the effective Hamiltonians derived from both the Berry phase and the weak drive limit. The linear term (in $\hat S_z$) is not present since we are in the corresponding rotating frame defined by $\tilde{\omega}_B$ already. We similarly find agreement with the effective dephasing derived via $N_J$ fluctuations in the End Matter and in the weak drive limit.

\section{Single-particle decoherence}
\label{sec:decoherence}

In this section, we derive the scaling behavior of the optimal squeezing and the corresponding squeezing time in the presence of single-particle decoherence. As in the main text, we focus on the cases of spontaneous emission and single-particle dephasing separately. Here, we use a short-time, perturbative expansion of the time evolution of $\xi^2$ based on the exact solution of the dynamics \cite{Chu2021}.

Before proceeding, we first consider the behavior in the absence of decoherence. For an OAT Hamiltonian $-\check{\chi} \hat S_z^2$, we find
\begin{equation}
    \xi^2(t) \approx \frac{1}{N^2 \check{\chi}^2 t^2} + \frac{N^2 \check{\chi}^4 t^4}{6}.
\end{equation}
Minimizing this in time, we find
\begin{equation}
    \xi^2_{\text{opt}} = \frac{3^{2/3}}{2 N^{2/3}}, \qquad
    \check{\chi} t_{\text{opt}} = \frac{3^{1/6}}{N^{2/3}},
\end{equation}
accurately capturing the usual scaling for OAT dynamics. 

Beyond OAT dynamics, our protocol also undergoes collective decoherence. For collective decoherence (jump operator $\hat{S}_z$) at rate $\check{\Gamma}$, we find \cite{Schleier-Smith2010,Lewis-Swan2018,Chu2021,Baamara2022}
\begin{equation}
    \xi^2(t) \approx \frac{1 + N \check{\Gamma} t}{N^2 \check{\chi}^2 t^2} + \frac{N^2 \check{\chi}^4 t^4}{6}.
\end{equation}
Assuming $N \check \Gamma t_{\text{opt}} \gg 1$, we optimize the squeezing, finding
\begin{equation}
    \xi^2_{\text{opt}} \approx \frac{5 (\check{\Gamma}/\check{\chi})^{4/5}}{2^{9/5}\times 3^{1/5}N^{2/5}}, \qquad t_{\text{opt}} \approx \frac{(3/2)^{1/5}(\check \Gamma/ \check \chi)^{1/5}}{\check \chi N^{3/5}},
\end{equation}
so the scaling depends on how we scale $\check \Gamma/\check \chi$. 

As discussed in Sec.~\ref{sec:HP}, $\check \Gamma/\check \chi$ is minimized for $2 \delta \sec \tilde{\theta}_J \ll f_J N \Gamma \cos \tilde{\theta}_J$, thus allowing for the greatest amount of squeezing. Working in this limit and taking $f_J = f_{\uparrow} = 1/2$, corresponding to the equator in the $\hat S$ Bloch sphere, we find
\begin{equation}
    \check{\chi} = \frac{ \delta \sin^2 \tilde{\theta}_J}{2 N \cos^3 \tilde{\theta}_J}, \qquad \check{\Gamma} = \frac{4 \delta^2 \sin^2 \tilde{\theta}_J}{N^2 \Gamma^2 \cos^6 \tilde{\theta}_J} \Gamma, \qquad \frac{\check{\Gamma}}{\check{\chi}} = \frac{8 \delta}{N \Gamma \cos^3 \tilde{\theta}_J}.
\end{equation}
If we fix $\check \Gamma/\check \chi \equiv a N^\beta$, then we find
\begin{equation}
    \xi^2_{\text{opt}} \approx \frac{ 5 a^{4/5}}{2^{9/5} \times 3^{1/5} N^{\alpha}}, \qquad \Gamma t_{\text{opt}} \approx \frac{16 (3/2)^{1/5}}{a^{4/5} \sin^2 \tilde{\theta}_J N^{1-\alpha}},
\end{equation}
where we have defined $\alpha \equiv (2-4 \beta)/5$. Here, we see explicitly the tradeoff in varying both $\alpha$ and $a$, with $\xi^2_{\text{opt}} \Gamma t_{\text{opt}} \sin^2 \tilde{\theta}_J \approx 20/N$, independent of $a$, as well as the role of $\sin^2 \tilde{\theta}_J$ in the squeezing timescale. Furthermore, we note that it is only sensible to consider $2/3 \geq \alpha > 0$, corresponding to $ -1/3 \leq \beta \leq 1/2$, as we cannot achieve better than OAT scaling and we do not consider non-scalable squeezing. Since we can always ensure $\check \Gamma/\check \chi \to 0$ by taking $\delta \to 0$, then any of these scaling behaviors can be achieved with the caveat that the total squeezing is always constrained by OAT squeezing, which limits the choice of $a$ for a given $N, \alpha$, and that we satisfy the approximations used to derive the squeezing dynamics (adiabatic elimination of $\hat \jmath$ and relevance of higher-order terms near criticality). 

\subsection{Spontaneous emission}
\label{sec:SE}

First, we consider the optimal squeezing in the presence of (effective) spontaneous emission in the $\hat S$ Bloch sphere. As discussed in the main text, the optimal squeezing occurs in the limit of weak drive. In light of this, we will utilize the OAT and collective dephasing rates derived in Sec.~\ref{sec:weak}. 
\begin{equation}
    \check{\chi} = \frac{\delta \Omega^2 (\delta \chi/2 + N \chi^2/4 + N \Gamma_\Delta^2/16)}{2 |\delta + (\chi + i \Gamma_\Delta/2)N/2|^4}, \qquad \check{\Gamma_\Delta} = \frac{\delta^2 \Omega^2 \Gamma_\Delta}{4 |\delta + (\chi + i \Gamma_\Delta/2)N/2|^4}.
\end{equation}
In the weak drive limit, we further find
\begin{equation}
    \gamma_- = \gamma_{e \uparrow} \sin^2 \frac{\tilde{\theta}_J}{2} \approx \frac{\Omega^2/4}{|\delta + (\chi + i \Gamma_\Delta/2)N/2|^2} \gamma_{e \uparrow}.
\end{equation}
To simplify the analysis, we will focus on two limits $\delta^2 \gg N^2 (\chi^2 + \Gamma_\Delta^2/4)/4$ and $\delta^2 \ll N^2 (\chi^2 + \Gamma_\Delta^2/4)/4$.

The limit $\delta^2 \gg N^2 (\chi^2 + \Gamma_\Delta^2/4)/4$ corresponds to far off-resonant drive. In this limit, we have
\begin{equation}
    \check{\chi} \approx \chi \sin^2 \frac{\tilde{\theta}_J}{2}, \qquad \check{\Gamma} \approx \Gamma_\Delta \sin^2 \frac{\tilde{\theta}_J}{2},
\end{equation}
and so the OAT rate is just the original two-level rate normalized by the excited state population while the collective decoherence rate is similarly the two-level collective dissipation rate renormalized by the excited state population. The detuning only enters by determining this excited state population. Defining $\tau \equiv t \sin^2 \frac{\tilde{\theta}_J}{2}$, we find
\begin{equation}
    \xi^2(\tau) \approx \frac{1 + N \Gamma_\Delta \tau}{N^2 \chi^2 \tau^2} + \frac{N^2 \chi^4 \tau^4}{6} + \frac{2}{3} \gamma_{e \uparrow} \tau.
\end{equation}
Optimizing the squeezing with respect to $\Delta$ and $\tau$, we find
\begin{equation}
    \xi^2_{\text{opt}} \approx \frac{3^{2/3}}{2 N^{2/3}} + \frac{4 \sqrt{2/3}}{\sqrt{N \Gamma/\gamma_{e \uparrow}}} , \qquad \gamma_{e \uparrow} t_{\text{opt}} \approx \frac{\sqrt{6}}{ \sin^2 \frac{\tilde{\theta}_J}{2}  \sqrt{ N \Gamma /\gamma_{e \uparrow}}}, \qquad \Delta_{ \text{opt}} \approx \pm \frac{\sqrt{\Gamma/\gamma_{e \uparrow}} 3^{1/3}N^{1/6}}{ \sqrt{2}} \frac{\kappa}{2},
\end{equation}
where we assume we are in a limit where $\Delta^2 \gg \kappa^2/4$ (which requires $ N^{1/3} \gg \gamma_{e \uparrow}/\Gamma$) for the optimization and note that the optimized values are expressed in terms of $\Gamma$ rather than $\Gamma_\Delta$. The squeezing will be reduced when this limit is not realized.
Logarithmic corrections can be obtained by going beyond linear approximations (in time) of the sources of decoherence. We note that the scaling of $\Delta_{\text{opt}}$ with $N$ can be increased to $\Delta_{\text{opt}} \sim (N \Gamma/\gamma_{e \uparrow})^{1/4} \kappa$ without (strongly) modifying the optimal scaling of the squeezing and squeezing time. Increasing the detuning modifies the scaling of the unitary OAT component of $\xi^2_{\text{opt}}$ from $N^{-2/3}$ to $N^{-1/2}$, so it at most modifies the coefficient of the dominant term. This can be utilized to ensure that $\Delta \gg \kappa$ is achieved.

The limit $\delta^2 \ll N^2 (\chi^2 + \Gamma^2/4)/4$ corresponds to near-resonant drive as in the main text and is described by
\begin{equation}
    \check{\chi} \approx  \frac{2 \delta }{N} \sin^2 \frac{\tilde{\theta}_J}{2}, \qquad \check{\Gamma} \approx \frac{\delta^2 \Gamma_\Delta}{N^2 \chi^2/4+N^2 \Gamma_\Delta^2/16} \sin^2 \frac{\tilde{\theta}_J}{2} = \frac{ 4 \delta^2 \kappa}{N^2 g_c^2} \sin^2 \frac{\tilde{\theta}_J}{2},
\end{equation}
where again the excited state population determines the overall time scale of the dynamics and we absorb it via $\tau$. 
Interestingly, we see that $\Delta$ does not play a role in this limit beyond its effect on the excited state population, indicating that there is no benefit (or detriment) from having elastic interactions. We can understand this as a consequence of the fact that, as we note in the End Matter, the Berry phase depends only on $\cos \tilde{\theta}_J$ and is otherwise independent of $\chi$. Optimizing the squeezing with respect to time and $\delta$, we find
\begin{equation}
    \xi^2_{\text{opt}} \approx  \frac{3^{2/3}}{2 N^{2/3}} + \frac{4 \sqrt{2/3}}{\sqrt{N \Gamma/\gamma_{e \uparrow}}} , \qquad \gamma_{e \uparrow} t_{\text{opt}} \approx \frac{\sqrt{6}}{ \sin^2 \frac{\tilde{\theta}_J}{2} \sqrt{N \Gamma/ \gamma_{e \uparrow}}}, \qquad \delta_{\text{opt}} \approx \pm \frac{\sqrt{\Gamma/\gamma_{e \uparrow}} N^{5/6}}{3^{1/3} \sqrt{2} } \frac{\gamma_{e \uparrow}}{2},
\end{equation}
where we can see that the scaling of $\delta$ is consistent with the limit we have considered and again we have expressed the results in terms of $\Gamma$ rather than $\Gamma_\Delta$. Logarithmic corrections can again be obtained by going beyond linear approximations (in time) of the sources of decoherence. Like for the dispersive regime, we may also modify the scaling of $\delta_{\text{opt}}$ to $\delta_{\text{opt}} \sim (N \Gamma/\gamma_{e \uparrow})^{3/4} \gamma_{e \uparrow}$ without changing the dominant scaling terms in the squeezing and squeezing times beyond a change to its coefficient. In contrast to the dispersive regime, here the relevant detuning becomes smaller. Remarkably, we also find the exact same optimal squeezing and squeezing time in both limits, implying that our protocol can potentially be extended to approaches where a weak drive and far off-resonant cavity were assumed, allowing them to go beyond the weak drive limit as well, subject to concerns of metastability in certain regions of the phase diagram \cite{Barberena2019}.

\subsection{Single-particle dephasing}

Next, we consider the optimal squeezing in the presence of single-particle dephasing. Unlike for spontaneous emission, we assume that the corresponding rate $\gamma_d$ is independent of the population. In this case, the squeezing dynamics is approximately
\begin{equation}
    \xi^2(t) \approx e^{\gamma_d t}\left(\frac{e^{\gamma_d t} + N \check{\Gamma} t}{N^2 \check{\chi}^2 t^2} + \frac{N^2 \check{\chi}^4 t^4}{6} \right).
\end{equation}
For $f_J = f_{\uparrow} = 1/2$ in the limit $\delta \ll N \Gamma \cos^2 \tilde{\theta}_J$ where optimal squeezing occurs, we have
\begin{equation}
    \check{\chi} = \frac{ \delta \sin^2 \tilde{\theta}_J}{2 N \cos^3 \tilde{\theta}_J}, \qquad \check{\Gamma} = \frac{4 \delta^2 \sin^2 \tilde{\theta}_J}{N^2 \Gamma^2 \cos^6 \tilde{\theta}_J} \Gamma, \qquad \frac{\check{\Gamma}}{\check{\chi}} = \frac{8 \delta}{N \Gamma \cos^3 \tilde{\theta}_J}.
\end{equation}
We see that both the OAT and the collective dephasing scale with $\sin^2 \tilde{\theta}_J$, indicating faster squeezing rates, and thus improved squeezing, can be obtained closer to criticality, although there are diminishing returns. Depending on the strength of $\gamma_d$, we identify two different optimal squeezing behaviors. For weak $\gamma_d \ll (3 e)^{2/3} N^{1/3} \sin^2 \tilde{\theta}_J \Gamma/32$, we find
\begin{equation}
    \xi^2_{\text{opt}} \approx \frac{3^{2/3}}{2 N^{2/3}} + \frac{4 \sqrt{10 \gamma_d/\Gamma}}{3^{1/6} N^{5/6} \sin \tilde{\theta}_J}, \qquad \gamma_d t_{\text{opt}} \approx \frac{ 8 \times 3^{1/6}}{\sqrt{10 \Gamma /\gamma_d} N^{1/6} \sin \tilde{\theta}_J }, \qquad \delta_{\text{opt}} \approx \pm \sqrt{\frac{5 N \Gamma \gamma_d}{8}} \frac{\cos^3 \tilde{\theta}_J}{\sin \tilde{\theta}_J} ,
\end{equation}
corresponding to ideal OAT squeezing, albeit at a slower rate. Here, we see that the optimal squeezing does not depend strongly on $\tilde{\theta}_J$, although the time scale and optimal detuning do.

For strong $\gamma_d \gg (3 e)^{2/3} N^{1/3} \sin^2 \tilde{\theta}_J \Gamma/32$, we find
\begin{equation}
    \xi^2_{\text{opt}} \approx \frac{3^{2/3} }{2 N^{2/3}} e^{4/3} + \frac{16 e \gamma_d}{N \Gamma \sin^2 \tilde{\theta}_J} , \qquad \gamma_d t_{\text{opt}} = 1, \qquad \delta_{\text{opt}} \approx \pm \frac{2 \times (3 e)^{1/6} N^{1/3} \cos^3 \tilde{\theta}_J }{\sin^2 \tilde{\theta}_J} \gamma_d,
\end{equation}
which will exhibit OAT scaling with an additional prefactor of $e^{5/3}(1+f_d)$, where $\gamma_d \equiv f_d \times (3 e)^{2/3} N^{1/3} \sin^2 \tilde{\theta}_J \Gamma/32$. Here, we see that the optimal squeezing time is simply the inverse of the dephasing rate, while the optimal squeezing is weakly dependent on $\sin^2 \tilde{\theta}_J$. In the main text, the example of $N= 10^6, \gamma_d = 100 \Gamma$ is nearly in this limit, which is why we found a time scale that was largely fixed as we moved away from the critical point while the squeezing itself was reduced. 
 
\subsection{Competing single-particle decoherence}

Finally, we consider scenarios where both sources of single-particle decoherence are present. Depending on the relative strength of both sources and the particular parameters, there will be a competition in terms of which is most relevant. When spontaneous emission is most dominant, the optimal squeezing will occur at weak drives with a finite detuning and $\xi^2_{\text{opt}} \propto 1/\sqrt{N}$. In contrast, when single-particle dephasing is most dominant, the optimal squeezing will occur near criticality with small detuning and $\xi^2_{\text{opt}}$. Hence when both are equally relevant, the optimal drive and detuning will be between these two extremes. In this section, we will determine when one or the other is dominant. 

In order to proceed, we must consider spontaneous emission beyond the weak drive limit. Using the values of $\check \chi, \check \Gamma$ obtained from adiabatic elimination of the HP equations, we find
\begin{equation}
\begin{gathered}
    \xi^2_{\text{opt}} \approx \frac{3^{2/3}}{2 N^{2/3}} + \frac{8 \sqrt{(1-\cos \tilde{\theta}_J)}}{\sin \tilde{\theta}_J \sqrt{3 N \Gamma /\gamma_{e \uparrow}} }, \qquad \gamma_{e \uparrow} t_{\text{opt}} \approx \frac{4 \sqrt{3}}{  \sin \tilde{\theta}_J \sqrt{ (1+\cos \tilde{\theta}_J) N  \Gamma / \gamma_{e \uparrow}}}, \\
    \delta_{\text{opt}} \approx \pm \frac{\cos^3 \tilde{\theta}_J \sqrt{(1- \cos \tilde{\theta}_J)  \Gamma/ \gamma_{e \uparrow}} N^{5/6}}{3^{1/3} \sin \tilde{\theta}_J} \frac{\gamma_{e \uparrow}}{2}
\end{gathered}
\end{equation}
We find that when $\cos \tilde{\theta}_J \to 1$, the squeezing is minimized and we recover the weak drive behavior, as anticipated. 

To determine the competition between both sources of decoherence, we determine where these optimal squeezing values (for fixed $\tilde{\theta}_J$) are the same for both spontaneous emission and for dephasing, indicating that the modification to the squeezing is the same for both decoherence processes. A more sophisticated approach might involve re-optimizing $\xi^2$ with the shift from the single-particle decoherence-free protocol doubled, although our less sophisticated approach will be sufficient in determining when one or the other is dominant, and thus whether we consider weak drive, critical drive, or somewhere in the middle.

For $f_d \ll 1$, the optimal squeezing values for the two scenarios are equal when
\begin{equation}
     \cos \tilde{\theta}_J \approx 1- \frac{15 }{2 \times 3^{1/3}  N^{2/3}} \frac{\gamma_d}{\gamma_{e \uparrow}},
\end{equation}
This equation is only valid for $0 \leq \cos \tilde{\theta}_J \leq 1$, so when $\gamma_d/\gamma_{e \uparrow} \ll 2 \times 3^{1/3} N^{2/3}/15$, we expect spontaneous emission to dominate and the optimal squeezing to occur at weak drive. In the opposite limit, we expect single-particle dephasing to dominate and the optimal squeezing to occur near criticality. This further requires $ \Gamma/\gamma_{e \uparrow} \sin^2 \tilde{\theta}_J \gg 64 (N/(3e)^2)^{1/3}/15$ to ensure $f_d \ll 1$. For the parameters in the main text, this requires $\Gamma/\gamma_{e \uparrow} \sin^2 \tilde{\theta}_J \gg 105$, which is not realized. 

For $f_d \gg 1$, the optimal squeezing values for the two scenarios are equal when
\begin{equation}
\label{eq:optdecse}
    \sin^2 \frac{\tilde{\theta}_J}{2} \cos \frac{\tilde{\theta}_J}{2} \approx \frac{2 e }{\sqrt{2 N \Gamma/\gamma_{e \uparrow}}} \frac{\gamma_d}{\gamma_{e \uparrow}},
\end{equation}
where we have utilized the fact that the OAT scaling terms are small for $f_d \gg 1$. Since the left-hand side varies from $2/(3 \sqrt{3})$ at criticality to 0 at weak drive, for $\gamma_d/\gamma_{e \uparrow} \gg  \sqrt{ N \Gamma/\gamma_{e \uparrow}}/(3 \sqrt{6} e)$, we predict single-particle decoherence to dominate and the optimal squeezing to occur near criticality. In the opposite limit $\gamma_d/\gamma_{e \uparrow} \ll \sqrt{ N \Gamma/\gamma_{e \uparrow}}/(3 \sqrt{6} e)$, we predict spontaneous emission to dominate and the optimal squeezing to occur at weak drive. When $\gamma_d/\gamma_{e \uparrow} \sim \sqrt{ N \Gamma/\gamma_{e \uparrow}}/(3 \sqrt{6} e)$, we predict both single-particle decoherence terms to be comparably relevant, and the optimal parameters will occur somewhere between the two extremes. For $\gamma_{e \uparrow} = \Gamma, N = 10^6$ as in the main text, this corresponds to $\gamma_d/\gamma_{e \uparrow} \sim 50$, hence the optimal drive occurring between the two extremes when $\gamma_d = 100 \Gamma$. Similar scaling emerges when comparing the optimal squeezing times or detunings, which both give the same result as one another. Hence we can also look at where the optimal detunings intersect for the two individual sources of single-particle decoherence, although this intersection will only be a very approximate estimate of the actual optimal parameters. 

\subsection{Comparison to other protocols}

In this section, we compare the scaling of the squeezing and squeezing time to other common approaches to the generation of spin squeezing in optical cavities. We focus only on the case of intrinsic sources of decoherence due to spontaneous emission, i.e., $\gamma_{e \downarrow} \sin^2 \frac{\tilde{\theta}_J}{2}$. However, as discussed in the main text, the ability of our protocol to go beyond the weak excitation regime and accelerate the squeezing dynamics provides a means of mitigating the effect of external dephasing sources when the squeezing is to be generated directly on the clock transition, as in the example level scheme of the main text. While the squeezing dynamics in all cases can be accelerated by generating it on transitions which couple more strongly to the cavity, this introduces the additional complication of transferring the spin squeezing to a clock transition, thereby introducing additional sources of noise. In all cases, we neglect overall timescales while keeping this fact in mind. We likewise neglect most constant coefficients for the optimal $\xi^2_{\text{opt}}$, which depend on the precise implementation and are typically comparable for all protocols. 

The protocols we consider are as follows:
\begin{itemize}
    \item (OAT) One-axis twisting $\hat H = \hat S_z^2$. We compare our protocol to the related protocol in the dispersive, weak excitation regime, see Sec.~\ref{sec:SE} or Ref.~\cite{Barberena2023b} for details about the squeezing optimization.
    \item (QND) Quantum non-demolition measurement utilizes a measurement jump operator $\hat l = \hat S_z$. In this approach, the emitted cavity light from the jump operator $\hat l$ is measured, providing a non-demolition measurement of $\hat S_z$, which reduces the variance in $\hat S_z$ at the cost of quantum backaction in $\langle \hat S_z \rangle$. Unlike the other protocols, this approach is non-unitary, and its performance depends strongly on the detection efficiency $\eta$. See Ref.~\cite{Barberena2023b} for details about the protocol and squeezing optimization as well as a discussion of the minimal $\eta$ necessary for QND to outperform OAT in the dispersive, weak-excitation regime.
    \item (TAT) Two-axis twisting $\hat H = \hat S_z^2 - \hat S_x^2$. In this approach, an additional counter-twisting term is introduced. In the absence of decoherence, TAT realizes Heisenberg squeezing $\xi^2 \sim 1/N$ along with accelerated squeezing dynamics due to the generation of an unstable saddle-point in the mean-field dynamics of the Bloch vector. See Ref.~\cite{Borregaard2017a} for details about the protocol and squeezing optimization.
    \item (TnT) Twist-and-turn $\hat H = \hat S_z^2 + \Omega \hat S_x$. In this approach, an additional drive term is added to mimic the instability like in TAT and accelerate the squeezing dynamics, although Heisenberg scaling is not realized. See Refs.~\cite{Muessel2015,Hu2017,Barberena2022} for details about the protocol and squeezing optimization.
\end{itemize}

In Table \ref{tab:protocols}, we summarize the scaling behavior of the spin squeezing and squeezing times in the presence of intrinsic sources of decoherence due to spontaneous emission. We note that like for our protocol, OAT and QND both exhibit different scaling depending on the effective form of the decoherence. This is a consequence of the fact that single-particle dephasing commutes with the associated dynamics and, more importantly, primarily modifies the anti-squeezed quadrature, leading to a reduced impact on the spin squeezing. For both TAT and TnT, this no longer holds, and both effective spin flips and dephasing impact the squeezing in similar ways. We also report several squeezing values realized experimentally in optical cavities.

\begin{table}[h]
\centering
\def\arraystretch{1}
\setlength\tabcolsep{2.4mm}
\begin{tabular}{ccccc}
\toprule
\toprule
 & \multirow{2}{*}{Decoherence} & \multirow{2}{*}{$\xi^2_{\text{opt}}$} & \multirow{2}{*}{$\gamma_e t_{\text{opt}}$} & Experiment \\
 & & & & ($N$ , $\xi^2$) \\
\midrule
 \multirow{2}{*}[-0.5\dimexpr \aboverulesep + \belowrulesep + \cmidrulewidth]{\shortstack{Our \\ Protocol}} & Spin Flips & $\frac{1}{\sqrt{NC}}$ & $\frac{1}{\sqrt{NC}}$ & (1000, $-8.2$ dB) [Simulation]\\
\cmidrule{2-4}
& Dephasing & $\frac{1}{N^{2/3}}$ & $\frac{1}{N^{2/3}}$ & (1000, $-14$ dB) [Simulation] \\
\midrule
 \multirow{2}{*}[-0.5\dimexpr \aboverulesep + \belowrulesep + \cmidrulewidth]{OAT} & Spin Flips & $\frac{1}{\sqrt{NC}}$ & $\frac{1}{\sqrt{NC}}$ & 
 \multirow{2}{*}[-0.5\dimexpr \aboverulesep + \belowrulesep + \cmidrulewidth]{($5 \times 10^4, -5.6$ dB) \cite{Leroux2010}, \hspace{3.2cm}} \\
\cmidrule{2-4}
& Dephasing & $\frac{1}{N^{2/3}}$ & $\frac{1}{N^{2/3}}$ & \hspace{3.2cm} ($1500, -6.5$ dB) \cite{Braverman2019} \\
\midrule
 \multirow{2}{*}[-0.5\dimexpr \aboverulesep + \belowrulesep + \cmidrulewidth]{QND} & Spin Flips & $\frac{1}{\sqrt{NC}}$ & $\frac{1}{\sqrt{NC}}$ & \multirow{2}{*}[-0.5\dimexpr \aboverulesep + \belowrulesep + \cmidrulewidth]{($5 \times 10^4$, $-3$ dB) \cite{Schleier-Smith2010a}, ($5 \times 10^5, -20$ dB) \cite{Hosten2016}, \hspace{3.4cm}} \\
\cmidrule{2-4}
& Dephasing & $\frac{e}{\eta N}\frac{C}{1+C}$ & $\frac{1}{1+C}$ & \hspace{6.8cm} ($5 \times 10^4$, $-18$ dB) \cite{Cox2016}\\
\midrule
TAT & Either & $\frac{1}{\sqrt{NC}}$ & $\frac{\log \sqrt{NC}}{\sqrt{NC}}$ & Recently implemented without measured squeezing in Ref.~\cite{Luo2024}.\\
\midrule
TnT & Either & $\frac{1}{\sqrt{NC}}$ & $\frac{\log \sqrt{NC}}{\sqrt{NC}}$ & (200, $-2$/$-4$ dB) \cite{Li2023a}* \\
 \bottomrule
 \bottomrule
\end{tabular}
\caption{Comparison of typical squeezing protocols for optical cavities and associated experimental realizations. $C \equiv 4 g_c^2/\kappa \gamma_e$ is the single-atom cooperativity, where $\gamma_e$ is the spontaneous emission rate of $|e\rangle$, and $\eta$ is the detection efficiency of the emitted cavity light. Experimental squeezing entries split between two rows exhibit both forms of single-particle decoherence. *Spin squeezing was measured for only a single suboptimal (in time) point; $-4$ dB represents the expected squeezing based on Ref.~\cite{Li2023a}'s numerical simulations, which show good agreement with the experimental data. \label{tab:protocols}}
\end{table}

Finally, we remark that our approach could be naturally extended to incorporate QND by measuring the emitted cavity light to weakly measure $\hat S_z$, whereby an analysis similar to Ref.~\cite{Barberena2023b}, which considered the weak excitation regime, could be implemented to determine when OAT- or QND-dominated squeezing is optimal for our approach.

\bibliography{StrongSqueezing,Extra}